\documentclass[aps,amsmath,amssymb,nofootinbib,superscriptaddress,showpacs,floatfix,prb,twocolumn]{revtex4-1}

\pdfoutput=1

\usepackage{latexsym}
\usepackage{graphicx}
\usepackage{times,psfrag,subfigure}
\usepackage{amsmath}
\usepackage{dcolumn}
\usepackage{bm, bbm}      
\usepackage{color}
\usepackage{latexsym,amsmath,amssymb,bm,euscript}
\usepackage[sort&compress]{natbib}
\def\Pf{\mathop{\mathrm{Pf}}}
\def\Tr{\mathop{\mathrm{Tr}}}
\def\sgn{\mathop{\textrm{sgn}}}

\newcommand{\beq}{\begin{equation}}
\newcommand{\eeq}{\end{equation}}
\newcommand{\beqarray}{\begin{eqnarray}}
\newcommand{\eeqarray}{\end{eqnarray}}
\newcommand{\Ref}[1]{Ref.~\onlinecite{#1}} 
\newcommand{\Sec}[1]{Sec.~\ref{#1}} 
\newcommand{\eq}[1]{Eq.~(\ref{#1})} 
\newcommand{\fig}[1]{Fig.~\ref{#1}} 
\newcommand{\figs}[1]{Figs.~\ref{#1}} 

\begin{document}

\allowdisplaybreaks

\title{Types of topological surface states in nodal noncentrosymmetric
  superconductors}

\date{\today}

\author{Andreas P. Schnyder}
\email{a.schnyder@fkf.mpg.de}
\affiliation{Max-Planck-Institut f\"ur Festk\"orperforschung,
  Heisenbergstrasse 1, D-70569 Stuttgart, Germany} 

\author{P. M. R. Brydon}
\email{brydon@theory.phy.tu-dresden.de}
\affiliation{Institut f\"ur Theoretische Physik, Technische Universit\"at
  Dresden, D-01062 Dresden, Germany}

\author{Carsten Timm}
\email{carsten.timm@tu-dresden.de}
\affiliation{Institut f\"ur Theoretische Physik, Technische Universit\"at
  Dresden, D-01062 Dresden, Germany}

\begin{abstract}
Nodal noncentrosymmetric superconductors have topologically nontrivial
properties manifested by protected zero-energy surface states.  
Specifically, it was recently found that zero-energy surface flat bands of
topological origin appear at their surface. 
We show that the presence of 
certain inversion-type  lattice symmetries can give rise to additional
topological features of the gap nodes, resulting in surface states forming
one-dimensional arcs connecting the projections of two nodal rings.  
In addition, we demonstrate that Majorana surface states can appear at
time-reversal-invariant momenta of the surface Brillouin zone, even when 
the system is not fully gapped in the bulk. 
Within a continuum theory we derive the
topological invariants that protect these different types of  zero-energy
surface states.  
We independently derive general conditions for the existence of zero-energy
surface bound states using the complementary
quasiclassical scattering theory, explicitly taking into account the effects of
spin-orbit splitting of the bands.
We compute surface bound-state spectra for various crystal point-group
symmetries and orbital-angular-momentum pairing states. Finally,
we examine the signatures of the arc surface states and of the zero-energy surface
flat bands in tunneling-conductance spectra and
dicuss how topological phase transitions in
noncentrosymmetric superconductors could be observed in experiments. 
\end{abstract}

\date{\today}



\maketitle

\section{Introduction}

Systems with strong spin-orbit coupling (SOC) have recently attracted a great
deal of attention. Prime examples are topological insulators,
where strong spin-orbit interactions give rise to a non-trivial band topology,
leading to topologically protected zero-energy surface
states.\cite{hasanKane2010,qiZhangReview2010,hasanMoore2011} Another class of
compounds for which SOC plays an important role are superconductors without
inversion symmetry.\cite{sigristIntro09,mineevSigristIntro09} In these
remarkable materials, Rashba-type antisymmetric spin-orbit interactions lift
the spin degeneracy of the electronic bands and generate complex spin textures
in the  electron Bloch functions.  In the superconducting state, the
antisymmetric  SOC gives rise to the admixture of even-parity spin-singlet and
odd-parity spin-triplet pairing components and, importantly, allows a
non-trivial topology of the Bogoliubov-quasiparticle
wavefunctions.\cite{satoPRB06,beriPRB2010,Schnyder2011,satoFujimoto2009} 
Akin to topological insulators, this non-trivial wavefunction topology results
in various types of protected zero-energy states at the edge or surface of
noncentrosymmetric superconductors 
(NCSs).\cite{Schnyder2011,satoFujimoto2009,Brydon2011,sato2011,yada2011,Tanaka2010,satoPRL10} 
For instance, a fully gapped NCS with nontrivial topology  supports
linearly dispersing helical Majorana modes at its
boundary.\cite{Schnyder2011,ryuNJP10,schnyder2008,SchnyderConfProc,royArxiv08,QiZhangPRL09,eschrigIniotakisReview,satoFujimoto2009}  
In three-dimensional systems, the stability of
these Majorana surface states is protected by an integer ($\mathbbm{Z}$)
topological invariant, i.e., the three-dimensional winding
number,\cite{Schnyder2011,schnyder2008}  whereas in two-dimensional systems a
binary ($\mathbbm{Z}_2$) topological number guarantees the robustness of the
edge
modes.\cite{ryuNJP10,satoFujimoto2009,satoPRB09,kaneMelePRL05b,QiHughesZhangPRB2010}   

Remarkably, topologically protected zero-energy boundary modes also occur in
NCSs with line
nodes.\cite{Schnyder2011,Brydon2011,sato2011,yada2011,Tanaka2010,satoPRL10}  
In particular, it has recently been shown that dispersionless zero-energy
states (i.e., flat bands) of topological origin generically appear at the
surface of three-dimensional nodal 
NCSs.\cite{Schnyder2011,Brydon2011}
These zero-energy  flat bands are confined to regions of the two-dimensional
surface Brillouin zone (BZ) that 
are bounded by the projections of the nodal lines of the bulk gap.\cite{foonoteNode}
The topological protection of these dispersionless boundary states is linked
to the  topological characteristics of the nodal gap structure via a
bulk-boundary correspondence. 
In fact, the stability of both the dispersionless zero-energy surface states
and the line nodes of the bulk gap is ensured by the conservation of the same
integer topological invariant, 
namely the one-dimensional winding number. Apart from these two-dimensional
surface flat bands, certain NCSs also support zero-energy boundary states that
form one-dimensional open arcs in the surface BZ, connecting the projection of
two nodal
rings.\cite{Tanaka2009,Vorontsov2008,Vorontsov2008b,Lu2010,Iniotakis2007,eschrigIniotakisReview,Brydon2011}   
Moreover, it has recently been reported that Majorana surface states can occur
at time-reversal-invariant momenta of the surface
BZ,\cite{Schnyder2011,satoPRL10} 
even if the superconductor is not fully gapped in the bulk.

Topological surface states are generic features of NCSs whose spin-triplet
pairing component is at least as strong as the spin-singlet one. 
It is therefore quite reasonable to expect that these zero-energy boundary
states occur in the superconducting state of the heavy fermion compounds
CePt$_3$Si,\cite{bauer04} 
CeRhSi$_3$,\cite{kimura05} and CeIrSi$_{3}$,\cite{sugitani06} as well as in
Y$_2$C$_3$,~\cite{amano04} Li$_2$Pd$_x$Pt$_{3-x}$B,\cite{togano04,badica05,yuan2006}
and 
Mo$_3$Al$_2$C.\cite{bauer2010,karki2010}
All of these noncentrosymmetric materials show strong spin-orbit interactions,
with the spin-orbit band splitting far exceeding the superconducting
energy scale.\cite{mineevSigristIntro09}
For the study of surface phenomena in NCSs it is thus important to explicitly
account for the strong  SOC,~\cite{Tanaka2010,Tanaka2009,Yokoyama2005,borkjePRB06,linderPRB07} a subtlety that has been
overlooked in many previous
works.\cite{Iniotakis2007,eschrigIniotakisReview,Vorontsov2008,Vorontsov2008b,Brydon2011,Lu2010}

The main aim of this paper is to provide a comprehensive classification of the
topological features of three-dimensional nodal NCSs. In particular, we
demonstrate 
 that the presence of certain inversion-type symmetries  
gives rise to additional topological features of the gap line nodes. These
features are characterized by a two-dimensional $\mathbbm{Z}_2$
topological invariant and 
manifest themselves on the surface as  one-dimensional arcs of zero-energy
states, terminating at the projection of the nodal lines onto the surface BZ. 
We also examine the appearance of Majorana surface states at
time-reversal-invariant momenta of the surface BZ and show that the topological properties
of these 
linearly dispersing modes are described by a one-dimensional  $\mathbbm{Z}_2$
invariant. 
Using a continuum model of NCSs, we derive expressions for the $\mathbbm{Z}_2$
invariants  and the winding number that protect the Majorana modes, the arc
surface states, and the  surface flat bands, respectively.  

We illustrate these topological features by investigating the 
surface-bound-state spectrum and the tunneling conductance of NCSs using
quasiclassical scattering theory. 
Allowing for non-negligible SOC, we consider three different 
experimentally relevant crystal point-group symmetries and  explore the
effects of  
pairing with higher orbital angular momenta [e.g., ($d_{x^2-y^2}+f)$-wave
pairing].  
We find that higher-orbital-an\-gu\-lar-momentum pairing leads to additional
topologically 
stable line nodes in the bulk gap. Correspondingly, there appear extra surface
flat bands associated with these additional nodal lines. We show that the
surface flat bands and the arc states leave unique signatures in the
tunneling-conductance spectra. Finally, we investigate topological phase transitions in
NCSs, i.e., zero-temperature quantum phase transitions 
where the momentum-space topology of the quasiparticle spectrum changes
abruptly as a function of the singlet-to-triplet ratio in the pairing
amplitude.  We argue that anomalies in the
density of states and the tunneling conductance  provide
experimental fingerprints of these zero-temperature phase transitions.

The remainder of this paper is organized as
follows. Section~\ref{sec:modelDef} discusses the model Hamiltonian and its
symmetries. In Sec.~\ref{secIII} we derive 
the topological invariants  characterizing  topological properties of both the
nodal lines and the surface states  
and give a detailed discussion of the topological criteria
for the existence of zero-energy surface states. By use of quasiclassical
scattering theory, we derive in Sec.~\ref{sec:boundStates}  general conditions 
for the occurrence of surface bound states in terms of sign changes of the
superconducting gap functions across the Fermi surface. We show that these
conditions are in perfect agreement with the topological criteria given in
Sec.~\ref{secIII}. In addition, we present in Sec.~\ref{sec:boundStates}
surface-bound state spectra for three different crystal point-group symmetries
and for various surface orientations. 
In Sec.~\ref{sec:tunnelingCond} we compute the tunneling conductance between a
normal metal and an NCS, 
identify the signatures of the zero-energy surface flat bands and the arc
surface states in the tunneling  spectra, and discuss topological phase
transitions. Our conclusions and outlook are given in
Sec.~\ref{sec:conclusions}.

\section{Model Hamiltonian and Symmetries}
\label{sec:modelDef}

We consider a three-dimensional single-band 
BCS superconductor with noncentrosymmetric crystal structure
and Rashba-type SOC. 
On a phenomenological level, such a superconductor is described by the
Bogoliubov-de Gennes Hamiltonian   
$\mathcal{H} = \frac{1}{2} \sum_{\bf{k}} \Psi^{\dag}_{\bf{k}} \mathcal{H} (
{\bf k} ) \Psi^{\ }_{\bf{k}}$, 
with
\begin{subequations} \label{def.ham}
\begin{eqnarray}
\mathcal{H} ( \bf{k} )
=
\begin{pmatrix}
h ( \bf{k} ) & \Delta ( \bf{k} ) \cr
\Delta^{\dag} ( \bf{k} ) & - h^{\textrm{T}} ( - \bf{k} )
\end{pmatrix} 
\end{eqnarray}
and $\Psi_{\bf{k} } = ( c^{\ }_{{\bf k} \uparrow}, c^{\ }_{\bf{k} \downarrow},
c^{\dag}_{- \bf{k} \uparrow}, c^{\dag}_{- \bf{k} \downarrow} )^{\textrm{T}}$, 
where $c^{\dag }_{\bf{k}\sigma}$ ($c^{\  }_{\bf{k}\sigma}$) denotes the
electron creation 
(annihilation) operator with momentum ${\bf k}$ and spin $\sigma$.
The normal-state dispersion of the electrons in the spin basis is given by 
\begin{eqnarray} \label{eq:NormState}
h ( {{\bf k}} )
&=&
\varepsilon_{{\bf k}} \sigma_0 + {\bf g}_{{\bf k}} \cdot \bm{\sigma },
\end{eqnarray}
where $\varepsilon_{{\bf k} } = \hbar^2 {\bf k}^2 / ( 2 m ) - \mu$,
${\bf g}_{{\bf k}}$  denotes the SOC potential,
${\bm{\sigma}} = (\sigma_x, \sigma_y, \sigma_z )^{\textrm{T}}$ are the three
Pauli matrices, and $\sigma_0$ stands for the $2 \times 2$ unit matrix.  
In the so-called helicity basis the normal-state Hamiltonian
\eqref{eq:NormState} takes diagonal form,  
$\tilde{h} ( {\bf k}) =  \mathrm{diag} ( \xi^+_{\bf k} , \xi^-_{\bf k} )$, where
$\xi^{\pm}_{\bf k} = \varepsilon_{\bf k} \pm \left| {\bf g}_{\bf k} \right|$ is
the dispersion of the positive-helicity and negative-helicity bands, respectively. 

Due to the lack of inversion symmetry, the superconducting gap generally contains
an admixture of  even-parity spin-singlet and odd-parity spin-triplet
pairing states,
\begin{eqnarray}
\Delta ( {{\bf k}} )
=
 \left(  \psi_{\bf k} \sigma_0 + {\bf d}_{{\bf k} } \cdot \bm{\sigma} \right)
\left( i \sigma_y \right) ,
\end{eqnarray}
\end{subequations}
where $\psi_{\bf k}$ and ${\bf d}_{\bf k}$ represent the spin-singlet and
spin-triplet components, respectively. 
It is well known that in the absence of
interband pairing, the superconducting transition temperature is maximized when 
the spin-triplet pairing vector ${\bf d}_{\bf k}$ is aligned with the
polarization vector ${\bf g}_{\bf k}$  
of the SOC.\cite{frigeriPRL04}
Hence, we parametrize the singlet and triplet components of the
superconducting gap function as 
\begin{subequations}  \label{gapParam}
\begin{eqnarray}
\psi_{\bf k}  
&=&	
\Delta_s f ( {\bf k} )
=
\frac{r}{r+1}\,  \Delta_0 f ({\bf k} ),	
\\
{\bf d}_{\bf k} 
&=&
\Delta_t f ( {\bf k} ) \, {\bf l}_{\bf k} 
=
\frac{1}{r+1}\,
\Delta_0 
f( {\bf k} ) \, {\bf l}_{\bf k} ,
\end{eqnarray}
\end{subequations}
where ${\bf l}_{\bf k} = {\bf g}_{\bf k} / \lambda$, with $\lambda$ the
SOC strength.  
Here, $r= \Delta_s / \Delta_t $ denotes the ratio between the singlet and
triplet pairing components. The pairing amplitudes $\Delta_s$ and
  $\Delta_t$ are  assumed to be positive and constant, i.e., $r\geq0$.
We have included in Eq.~\eqref{gapParam} a structure factor $f( {\bf k} $),
which allows us to investigate the effects of higher-orbital-angular-momentum 
pairing. In the following, we consider the three cases
\begin{eqnarray} 
f ({\bf k} )
= 
\left\{
\begin{array}{l l}
1 				& \quad \textrm{for ($s + p$)-wave}, \\
(k_x^2 - k_y^2)/k_F^2   	& \quad \textrm{for ($d_{x^2-y^2}+ f$)-wave},	 \\
2 k_x k_y / k_F^2 		& \quad \textrm{for ($d_{xy} + p$)-wave},
\end{array}
\right.
\end{eqnarray} 
where $k_F$ is the Fermi vector in the absence of spin-orbit interactions.
With Eq.~\eqref{gapParam}, it follows that the gaps on the two helicity bands
are given by 
$\Delta^{\pm}_{\bf k} = f ( {\bf k} )  \Delta_0 ( r \pm \left| {\bf l}_{\bf k}
\right| ) / (r+1)$. 

The specific form of the SOC vector ${\bf g}_{\bf k}$  (and
hence of ${\bf l}_{\bf k}$ and ${\bf d}_{\bf k}$) is constrained by
time-reversal symmetry and the  
crystallographic point-group symmetries of the superconductor.
Time-reversal symmetry requires ${\bf g}_{\bf k}$ to be real and an odd
function of ${\bf k}$. 
An element $R$ of the crystallographic point group $\mathcal{G}$ of the NCS
can be represented as either a proper (for $\det R = +1$) or an
improper (for $\det R = -1$) rotation. 
Thus, the Bogoliubov-de Gennes Hamiltonian~\eqref{def.ham} transforms under an
operation $R$ of $\mathcal{G}$ as
\begin{eqnarray} \label{rotTrafo}
U_{\tilde{R}} \mathcal{H} (  R^{-1}  {\bf k} ) U^{\dag}_{\tilde{R}} = +
\mathcal{H} ( {\bf k} )
\end{eqnarray}
with $\tilde{R} = \det (R) R$, $ U_{\tilde{R}} = \mathrm{diag} ( u_{\tilde{R}}
, u^{\ast}_{\tilde{R}} )$, and $u_{\tilde{R}}$ 
the spinor representation of $\tilde{R}$, i.e., 
$ u_{\tilde{R}} 
= \exp \left[ - i ( \theta /2 ) \hat{{\bf n}} \cdot \bm{\sigma}  \right] 
$.
Here,  $ \hat{{\bf n}} $ denotes the unit vector along the rotation axis of
$\tilde{R}$ and $\theta$ is the angle of rotation. 
It follows from Eq.~\eqref{def.ham} that the lattice symmetries
\eqref{rotTrafo} impose the constraint
\begin{eqnarray} \label{symG}
{\bf g}_{\bf k}  = \det (R) R \, {\bf g}_{R^{-1} {\bf k}} .
\end{eqnarray}
on ${\bf g}_{\bf k}$.
To determine the form of ${\bf g}_{\bf k}$, we employ a
small-momentum expansion. Using Eq.~\eqref{symG}  we find that for the
tetragonal point group  
$\mathcal{G} = C_{4v}$ (relevant for CePt$_3$Si, CeRhSi$_3$, and CeIrSi$_3$)
the lowest-order symmetry-allowed term is\cite{samokhin09}  
\begin{subequations} \label{Gkdef}
\begin{eqnarray} \label{SOC_C4vgroup}
{\bf g}_{\bf k}  
=
\lambda ( k_y {\bf \hat{x}} - k_x {\bf \hat{y}}) .
\end{eqnarray}
For the cubic point group $\mathcal{G} = O$ (represented by
Li$_2$Pd$_x$Pt$_{3-x}$B and Mo$_3$Al$_2$C)
the vector ${\bf g}_{\bf k}$ takes the form
\begin{eqnarray} \label{SOC_Ogroup}
{\bf g}_{\bf k}  
&=&
\lambda \Big[
k_x \left( 1 + g_2 \left[ k_y^2 + k_z^2 \right] \right) {\bf \hat{x}} 
+
k_y \left( 1 + g_2 \left[ k_x^2 + k_z^2 \right] \right) {\bf \hat{y}} 
\nonumber\\
&&
{} + k_z \left( 1 + g_2 \left[ k_x^2 + k_y^2 \right] \right) {\bf \hat{z}} 
\Big]
\end{eqnarray}
with the second-order SOC $g_2$.
Finally,  for the tetrahedral point group $\mathcal{G} = T_{d}$
(relevant for Y$_2$C$_3$) the small-momentum expansion 
of ${\bf g}_{\bf k}$ yields
\begin{equation}
{\bf g}_{\bf k}  
=
\lambda \Big[
k_x ( k_y^2 - k_z^2 ) {\bf \hat{x}} 
+
k_y ( k_z^2 - k_x^2 ) {\bf \hat{y}} 
+
k_z ( k_x^2 - k_y^2 ) {\bf \hat{z}} 
\Big].
\\
\end{equation} 
\end{subequations}

Before deriving the relevant topological invariants for
Hamiltonian~\eqref{def.ham}, we
first discuss in some detail the discrete symmetries responsible for the
nontrivial topological characteristics of $\mathcal{H} ({\bf k})$. 
According to the classification of Refs.~\onlinecite{schnyder2008},
\onlinecite{zirnbauer96}, and \onlinecite{altlandZirnb97}, 
$\mathcal{H} ( {\bf k} )$ belongs to symmetry class DIII
since it satisfies two independent antiunitary discrete symmetries:
Particle-hole symmetry (PHS) $\mathcal{C} = \mathcal{K} U_C$, with
\mbox{$U_C=\sigma_1 \otimes \sigma_0$},\cite{footnoteTensorProd} 
and time-reversal symmetry (TRS) $\mathcal{T} = \mathcal{K} U_T$, with $U_T =
\sigma_0 \otimes i \sigma_2$, where $\mathcal{K}$ denotes the complex
conjugation 
operator. TRS acts on the Bogoliubov-de Gennes Hamiltonian
$\mathcal{H} ( \bm{k} )$ as 
\begin{eqnarray} \label{eqTRS}
U_T \mathcal{H}^{\textrm{T} } (  -{ \bf k} ) U^{\dag}_T = + \mathcal{H} ( {\bf
  k} ) , 
\end{eqnarray}
and PHS as
\begin{eqnarray} \label{eqPHS}
U_C \mathcal{H}^{\textrm{T}} ( - {\bf k} ) U^{\dag}_C = - \mathcal{H} ( {\bf k} ) .
\end{eqnarray}
Combining TRS and PHS, one obtains a third
discrete symmetry, the so-called chiral symmetry, which acts as
\begin{eqnarray} \label{eqSLS}
U^{\dag}_S \mathcal{H} ({\bf k} ) U^{\phantom{\dag}}_S = - \mathcal{H} ( {\bf k} )
\end{eqnarray}
with the unitary matrix $U_S = i U_T U_C = - \sigma_1 \otimes \sigma_2$.
As we will explain in Sec.~\ref{secIII},
it is the chiral symmetry \eqref{eqSLS}  that leads
to the protection of the zero-energy surface flat bands.
Since $\mathcal{H} ( {\bf k} )$ anticommutes with the unitary matrix
$U_S$, it can be brought into block off-diagonal form. This is achieved by a
 unitary transformation that diagonalizes $U_S$, e.g.,  $W U_S W^{\dag}=
 \mathrm{diag} ( \sigma_0, -\sigma_0 )$,  
with
\begin{subequations} \label{offDiagTrafo}
\begin{eqnarray}
W
=
\frac{1}{\sqrt{2} }
\begin{pmatrix}
+  \sigma_0 & - \sigma_2 \cr
+ i \sigma_2 & + i\sigma_0 \cr
\end{pmatrix} .
\end{eqnarray}
The transformed Hamiltonian reads
\begin{eqnarray} \label{Htranf}
\tilde{ \mathcal{H}} ( {\bf k} )
=
W \mathcal{H} ( {\bf k} ) W^{\dag}
=
\begin{pmatrix}
 0 & D^{\phantom{\dag}}_{\bf k} \cr
 D^{\dag}_{\bf k} & 0 \cr
\end{pmatrix} ,
\end{eqnarray}
where the off-diagonal block is given by
\begin{eqnarray} \label{blockOffDk}
D_{\bf k}
=
 \left(
 B_{\bf k} \sigma_0  +  A_{\bf k}  \, {\bf l}_{\bf k} \cdot {\bm \sigma} 
\right) ( - i \sigma_2 )
\end{eqnarray}
\end{subequations}
with the short-hand notation $A_{\bf k} = \lambda + i \Delta_t f ( {\bf k} )$
and $B_{\bf k} = \varepsilon_{\bf k} + i \Delta_s f ( {\bf k} )$. 
We will see in the next section that the topological invariants
characterizing the  topology properties of $\mathcal{H} ( {\bf k})$ 
can be conveniently defined in terms of the off-diagonal block $D_{\bf k}$,
Eq.~\eqref{offDiagTrafo}, or its flat-band version. It is found that a
crystallographic point-group operation $R$ acts on $D_{\bf 
   k}$ as $u_{\tilde{R} } D_{  R^{-1} {\bf k}} u^{\textrm{T} }_{\tilde{R}} =
D_{ {\bf k} }$, whereas TRS implies  $D_{\bf k}  = -
D^{\textrm{T} }_{ - {\bf k} }$.

Depending on the point group $\mathcal{G}$ and the specific form of 
${\bf g}_{\bf k}$, the Hamiltonian $\mathcal{H} ({\bf {k}})$ may in addition
to~\eq{eqTRS} also satisfy another ``time-reversal''-like symmetry that acts only on
a two-dimensional plane within the three-dimensional BZ, i.e. a symmetry
that acts on $D_{\bf k}$ as
\begin{subequations}\label{newTRS}
\begin{eqnarray} \label{newTRSD}
 - D^{\textrm{T} }_{ - { \bf k}_i , k_0} = D_{{ \bf k}_i , k_0},
\end{eqnarray}
with ${\bf k}_i = ( k_{i1}, k_{i2} )$  the coordinates within the plane
$E_{k_0} ( \hat{{\bf u}} )  = \{{\bf k}  \, | \,  {\bf k} = k_0  \hat{\bf u}
+ k_{i1}  \hat{{\bf v}} + k_{i2}  \hat{{\bf w}} \}$ 
and $k_0$ the coordinate perpendicular to $E_{k_0}$.
Here, $\hat{{\bf u}} =  \hat{{\bf v}} \times  \hat{{\bf w}}$ and
$\hat{{\bf v}}$ and $\hat{{\bf w}}$ are taken to be orthogonal unit vectors.
It follows from Eq.~\eqref{blockOffDk} that a combination of the  symmetries
\eqref{eqTRS} and \eqref{newTRSD} imposes the constraint
\begin{eqnarray} \label{TRSgSec}
 {\bf g}_{ {\bf k}_i, - k_0} =    {\bf g}_{ {\bf k}_i, + k_0 }
\end{eqnarray}
\end{subequations}
on ${\bf g}_{\bf k}=\lambda{\bf l}_{\bf k}$. For example,
in the case of the tetragonal point-group
$C_{4v}$ with ${\bf g}_{\bf k}$ given by Eq.~\eqref{SOC_C4vgroup},  symmetry
\eqref{newTRS} is satisfied for the one-parameter family of planes $E_{k_z} ({
  \hat{\bf z}})$.
It is important to note that even though Eq.~\eqref{TRSgSec} has the form of
an inversion-type symmetry, it is different from the mirror symmetries imposed 
by the crystallographic point group, see Eq.~\eqref{symG}. We will explain
below that the presence of symmetry~\eqref{newTRS} can give rise to
topologically stable arc surface states.

\section{Topological invariants in nodal noncentrosymmetric superconductors}
\label{secIII}

To characterize the topological properties of nodal NCSs in three dimensions, we
introduce an integer topological invariant and two $\mathbbm{Z}_2$ topological
numbers. For this purpose it is convenient 
to work in the off-diagonal basis~\eqref{offDiagTrafo} and to adiabatically
deform $\tilde{\mathcal{H}} ({\bf k})$ into a flat-band Hamiltonian.  
The  off-diagonal block $D_{\bf k}$, which is in general non-Hermitian,
can through a singular-value decomposition be written as 
$D_{ \bf k  } = U^{\dag}_{ \bm{k} } \Sigma^{\phantom{\dag}}_{ \bm{k} }
V^{\phantom{\dag}}_{ \bm{k} }$, 
where $U_{\bf k}$ and $V_{\bf k}$ are unitary matrices and
$\Sigma^{\phantom{\dag}}_{ \bm{k} }$ is 
a diagonal matrix with the positive eigenenergies
\begin{subequations} \label{BCSspectrum}
\begin{eqnarray} 
\Lambda_{1,{\bf k}}& = & \sqrt{ ( \xi^{+}_{\bf k} )^2 + (\Delta^{+}_{\bf k}
  )^2 } , \\
\Lambda_{2,{\bf k}}& = & \sqrt{ ( \xi^-_{\bf k} )^2 + (\Delta^-_{\bf k} )^2 }
\end{eqnarray}
\end{subequations}
of $\tilde{\mathcal{H}} ( {\bf k} )$ and thus of $\mathcal{H} ( {\bf k} )$
on its diagonal, i.e., $\Sigma^{\phantom{\dag}}_{ \bm{k} }=
\textrm{diag} ( \Lambda_{1, {\bf k} } , \Lambda_{2 , {\bf k}} )$.
Assuming that $\Lambda_{ 1, {\bf k} }, \Lambda_{ 2, {\bf k} } \ne 0$, we
adiabatically deform the spectrum of $\tilde{\mathcal{H}} ( {\bf k})$ 
into flat bands with eigenvalues $+1$ and $-1$, which amounts
to replacing $\Sigma_{\bf k}$ by the unit matrix.
Hence, the off-diagonal component of the flat-band Hamiltonian is given
by\cite{Schnyder2011}	 
\begin{eqnarray} \label{qForNCS}
q ({ \bf k} ) & = &
 U^{\dag}_ {\bf k}  V^{\phantom{\dag}}_{\bf k} 
=
\frac{1}{2 \Lambda_{1,{\bf k}} \Lambda_{2, {\bf k}} }
\bigg[ 
\Big\{ 
\Lambda^+_{{\bf k}} B^{\phantom{+}}_{\bf k}
- \Lambda^-_{{\bf k}} A^{\phantom{+}}_{\bf k} \left| {\bf l}_{\bf k} \right| 
\Big\} \sigma_0
\nonumber\\
&& {}+ \Big\{
\Lambda^+_{{\bf k}} A^{\phantom{+}}_{\bf k} \left| {\bf  l}_{\bf k} \right|
-
\Lambda^-_{{\bf k}} B^{\phantom{+}}_{\bf k}
\Big\}  \frac{ {\bf l}_{\bf k} }{ \left| {\bf  l}_{\bf k} \right| } \cdot \bm{\sigma} 
\bigg]  ,
\end{eqnarray}
where $\Lambda^{\pm}_{\bf k} = \Lambda_{1, {\bf k}} \pm \Lambda_{2, {\bf k}}$.
As a result of TRS, the $2 \times 2$ unitary matrix $q (
{\bf k} ) \in U(2)$ satisfies 
$i \sigma_2 q^{\textrm{T}} ( {- \bf k} ) = q({  \bf k} ) i \sigma_2$.

\subsection{Winding number}

In a three-dimensional nodal superconductor, $q ( {\bf k})$ is defined only
for values of ${\bf k}$ for which
$\Lambda_{ 1, {\bf k} }, \Lambda_{ 2, {\bf k} } \ne 0$, i.e.,  for all ${\bf k}$
in the BZ except for those on the nodal lines.  
Hence, the three-dimensional winding number $\nu$,\cite{schnyder2008} which
characterizes fully gapped systems, is ill-defined for  
nodal NCSs. However, we can use the one-dimensional winding number
$W_{\mathcal{L}}$, 
which is defined in terms of a one-dimensional momentum-space loop (or line)
integral, to characterize the topology of nodal NCSs. 
To that end, we consider $q( {\bf k} )$ along a one-dimensional
loop (or line) in reciprocal space $\mathbbm{R}^3$ which does not cross
gapless regions.  
That is, we study the mapping  $S^1 \to U(2)$ given by $q ( {\bf k} ) \in
U(2)$. Since the first homotopy group of $U(2)$ is $\pi_1 [ U(2) ] =
\mathbbm{Z}$,\cite{nakahara} there is  
an infinite number of homotopy classes of mappings from $S^1$ to $U(2)$, which
can be labeled by the one-dimensional winding number 
\begin{eqnarray} \label{eq: 1D winding no} 
W_{\mathcal{L}} =
\frac{1}{2 \pi i} 
\oint_{ \mathcal{L} } d k_l \, \mathrm{Tr} 
\left[
q^{\dag} ( {\bf k} )  \partial_{k_l}  q^{ } (  {\bf k} )  
\right] .
\end{eqnarray}
The integral is to be evaluated along the path $\mathcal{L}$
parametrized by $k_l$. 
We observe that $W_{\mathcal{L}}$ is quantized for any closed loop
$\mathcal{L}$ that does not intersect with nodal lines. If
$\mathcal{L}$  encircles a line node, $W_\mathcal{L}$ determines the 
topological charge and hence the topological stability of the nodal line.
Using Eq.~\eqref{qForNCS} we find
\begin{eqnarray}
\lefteqn{ W_{\mathcal{L}} =
\frac{1}{2 \pi i } \oint_{\mathcal{L}} d k_l \, \partial_{k_l} 
\left[ \ln \det q ( {\bf k} ) \right] } \nonumber\\
&& = \frac{1}{2 \pi  } \oint_{\mathcal{L}} d k_l \, \partial_{k_l}
\left[ \arg \left( B^2_{\bf k} - A^2_{\bf k}  \left| {\bf l}_{\bf k} \right|^2  \right) \right] 
\nonumber \\
&& = \frac{1}{2 \pi  } \oint_{\mathcal{L}} d k_l \, \partial_{k_l}
\left[
\arg \left( \xi^+_{\bf k} + i \Delta^+_{\bf k} \right) + \arg \left( \xi^-_{\bf k} + i \Delta^-_{\bf k} \right) 
\right] ,
\nonumber \\
\end{eqnarray}
where, in going from the first to the second line, we have used the identity
$\left| \det q ({\bf k}) \right| = 1$.  

Assuming that the energy scales of the gap functions are much smaller than
those of the normal-state dispersions  
(i.e, for a weak-pairing superconductor), we can rescale $\Delta^{\pm}_{\bf
  k}$ without changing $W_{\mathcal{L}}$, such that the gaps are  
nonzero only within a small neighborhood of the Fermi surface sheets
[cf.~Eq.~\eqref{BCSspectrum}].
In this limit we find that 
the phase $\arg ( \xi^{\nu}_{\bf k} + i \Delta^{\nu}_{\bf k} )$ is constant
far away from the Fermi surface of helicity $\nu=\pm$ and that it jumps by  
\begin{eqnarray} \label{jumpTH}
 - \pi \sgn \big( \left. \partial_{k_l} \xi^{\nu}_{\bf k} \right|_{{\bf k}= {\bf k}^0_F }
   \big) \sgn \big( \Delta^{\nu}_{{\bf k}^0_F} \big)
\end{eqnarray}
where the path~$\mathcal{L}$ crosses the Fermi surface at ${\bf k}^0_F$.
It follows that the winding number $W_{\mathcal{L}}$ can be simplified to 
(cf.\ Ref.\ \onlinecite{sato2011})
\begin{eqnarray} \label{WLsimpI}
W_{\mathcal{L}}
=
- \frac{1}{2}
\sum_{\nu = \pm } \, \sum_{{\bf k}^0_F \in S^{\nu}_{\mathcal{L}} } 
 \sgn \big( \left. \partial_{k_l} \xi^{\nu}_{\bf k} \right|_{{\bf k}= {\bf k}^0_F }  \big)
  \sgn \big( \Delta^{\nu}_{{\bf k}^0_F} \big) , \; \; \; \;
\end{eqnarray}
where the set of points $S^{\nu}_{\mathcal{L}}$ is given by the intersection
of the path $\mathcal{L}$ with the Fermi surface for helicity $\nu$.
Thus Eq.~\eqref{WLsimpI} shows that for a weak-pairing NCS $W_{\mathcal{L}}$ is
completely determined by the phase structure of the superconducting gaps
$\Delta^+_{\bf k}$ and $\Delta^-_{\bf k}$ in the vicinity 
of the positive-helicity and negative-helicity Fermi surfaces, respectively.
Moreover, it follows from Eq.~\eqref{WLsimpI}  that nodal lines necessarily
carry nontrivial topological charge $W_{\mathcal{L}} \ne 0$, irrespective of
the particular form of the band structure or the crystal point-group
symmetries.  

\subsection{Two-dimensional $\mathbbm{Z}_2$ topological invariant}

The two-dimensional $\mathbbm{Z}_2$ topological invariant introduced in
this subsection is defined only for NCSs satisfying 
symmetry \eqref{newTRS}. We therefore consider a Hamiltonian $\mathcal{H}
( {\bf k})$ of the form~\eqref{def.ham} that is invariant under 
symmetry~\eqref{newTRS} for the plane $E_{k_0} ( \hat{\bf u})$. Furthermore,
we assume that the nodal lines of $\mathcal{H} ( {\bf k} )$ 
do not cross $E_{k_0} ( \hat{\bf u})$. Hence, $\mathcal{H} ( {\bf k} )$
restricted to the plane $E_{k_0} ( \hat{\bf u})$ 
can be regarded as describing a two-dimensional fully gapped superconductor invariant
under both TRS [i.e., symmetry \eqref{newTRS}] and 
PHS [i.e., the combination of chiral symmetry and symmetry
  \eqref{newTRS}].  
Such a two-dimensional system belongs to symmetry class DIII and its
topological characteristics  are described by the 
$\mathbbm{Z}_2$ topological
invariant\cite{ryuNJP10,satoFujimoto2009,satoPRB09,kaneMelePRL05b,QiHughesZhangPRB2010,Schnyder2011}   
\beqarray  
N^{\textrm{2D}}_{E_{k_0}} &
= 
&\prod_{k_{i2}=0, \pi}
\frac{ \Pf \left[  i \sigma_2 \, \hat{q}  ( \pi, k_{i2}, k_0 ) \right]  }
{ \Pf \left[  i \sigma_2 \, \hat{q}   ( 0 , k_{i2}, k_0 ) \right]  } \notag \\
&& {} \times
e^{ - \frac{1}{2} \int\limits_0^{\pi } d k_{i1} \Tr \left[ \hat{q}^{\dag} ({\bf k} )
  \, \partial_{k_{i1}}  \hat{q}  ( {\bf k} )   \right] }  , \label{Z2noLattice}
\eeqarray
where $\hat{q} ( {\bf k} )$ represents a lattice regularization 
of $q ( {\bf  k} )$\cite{footnoteLatticeReg}
  and $N^{\textrm{2D}}_{E_{k_0}} = -1$ ($+1$) indicates  
a topologically nontrivial (trivial) character. In Eq.~\eqref{Z2noLattice}, we 
have assumed that the coordinates $(k_{i1}, k_{i2})$ within the plane $E_{k_0}
( \hat{\bf u})$ are chosen such that ${\bf K}_i = (0,0)$, $(\pi, 0)$, $(0,
\pi)$, and $(\pi, \pi)$ are left invariant under symmetry~\eqref{newTRS}. Note
that at these points $i \sigma_2 \hat{q} ( {\bf K}_i, k_0 )$ is
antisymmetric, so that the Pfaffian is well-defined.

As before, we consider the weak-pairing limit and set the gaps
$\Delta^{\pm}_{\bf k}$ to zero far away from the Fermi surfaces.  
Using Eq.~\eqref{WLsimpI}, we find that in this approximation the exponential
factor in Eq.~\eqref{Z2noLattice} reduces  to  
\begin{eqnarray}
\prod_{\nu = \pm } \: \prod_{{\bf k}^0_F \in \tilde{S}^{\nu}_{E_{k_0}} }
i \sgn \big(
 \Delta^{\nu}_{{\bf k}^0_F} \left. \partial_{k_{i1}} \xi^{\nu}_{\bf k} \right|_{{\bf k}= {\bf k}^0_F }
 \big),
\end{eqnarray}
where the set of points $\tilde{S}^{\nu}_{E_{k_0}}$ is given by the
intersection of the Fermi surface for helicity $\nu$ with the integration paths
${\bf k}_i:  (0,0) \to (\pi, 0)$ and ${\bf k}_i:(0,\pi) \to (\pi, \pi)$. 
Assuming that the Fermi level does not cross the positive-helicity or negative-helicity
bands at $( {\bf K}_i, k_0)$, we find  that 
$\Pf [ i \sigma_2 \hat{q} ( {\bf K}_i, k_0 )  ]$ is either $+1$ or $-1$
depending on whether the helicity bands  at $( {\bf K}_i, k_0 )$ are occupied
or unoccupied. 
As a result, the $\mathbbm{Z}_2$ invariant for $\mathcal{H} ( {\bf k})$
simplifies to 
\begin{eqnarray} \label{Z2simp}
N^{\textrm{2D}}_{E_{k_0}} =
\sgn \big( \Delta^+_{E_{k_0}} \big) \sgn \big( \Delta^-_{E_{k_0}} \big) ,
\end{eqnarray}
where $\sgn ( \Delta^\pm_{E_{k_0}})$ denotes the sign of the gap on the Fermi
line given 
by the intersection of $E_{k_0} ( \hat{\bf u})$ with the
positive/negative-helicity Fermi surface. Observe that  Eq.~\eqref{Z2simp} does not depend on
the lattice regularization and hence is valid also in the continuum limit. 

For a NCS that is invariant under symmetry~\eqref{newTRS} for a
\emph{one-parameter family} of planes $E_{k} ( \hat{\bf u})$, 
with for example $k \in \mathbbm{R}$,
it is possible to assign a $\mathbbm{Z}_2$ topological
charge to the line nodes, provided that there are fully gapped
regions in momentum space separating different  line nodes.
For concreteness, let us consider the situation where a nodal line is located
within the plane $E_{k=k_0} ( \hat{\bf u})$. The $\mathbbm{Z}_2$ topological
charge of this nodal ring can be defined as
\beq \label{Z2ChArge}
\tilde{N}^{\textrm{2D}}
=
 N^{\textrm{2D}}_{E_{k^+_0}}  N^{\textrm{2D}}_{E_{k^-_0}} ,
\eeq
where $E_{k^+_0}$ and $E_{k^-_0}$  represent two planes that are located on
either side of the nodal line. For the tetragonal point-group $C_{4v}$ [with
  ${\bf g}_{\bf k}$ of the form \eqref{SOC_C4vgroup}], the two-dimensional
$\mathbbm{Z}_2$ number can be defined for the one-parameter family of planes
$E_{k_z} (\hat{\bf z} )$. The nodal rings, which are
centered around the $k_z$-axis, consequently carry a nontrivial
$\mathbbm{Z}_2$ topological charge $\tilde{N}^{\textrm{2D}}_{\alpha}  =
N^{\textrm{2D}}_{E_{k_z=0}}  N^{\textrm{2D}}_{E_{k_z = \alpha k_F}}$, 
where $k_F = \sqrt{ 2 m \mu / \hbar^2}$ and $\alpha=\pm$ distinguishes between
the topological charges of the nodal rings in the upper ($k_z > 0$) and  lower
($k_z < 0$) half-spaces.

\subsection{One-dimensional $\mathbbm{Z}_2$ topological invariant}

Finally, we also introduce a one-dimensional
$\mathbbm{Z}_2$  invariant that characterizes the topological properties of
$\mathcal{H} ({\bf k} )$  
restricted to a \emph{time-reversal-invariant} loop (or line) $\mathcal{L}$,
which is mapped onto itself under ${\bf k} \to -{\bf k}$. 
Similarly to Eq.~\eqref{Z2noLattice}, the one-dimensional $\mathbbm{Z}_2$
topological number can be conveniently defined in terms  
of the lattice-regularized version of $q ({\bf k})$,
Eq.~\eqref{qForNCS},\cite{kaneMelePRL05b,satoPRL10,QiHughesZhangPRB2010,Schnyder2011}   
\begin{eqnarray}  \label{1dZ2first}
N^{\textrm{1D}}_{\mathcal{L}}
=
\frac{ \Pf \left[  i \sigma_2 \, \hat{q}  ({\bf K}_2 ) \right]  }
{ \Pf \left[  i \sigma_2 \, \hat{q} ( {\bf K}_1 ) \right]  }
e^{ - \frac{1}{2} \oint_{\mathcal{L}} d k_{l} \Tr \left[ \hat{q}^{\dag} ( {\bf k} )
  \, \partial_{k_{l}}  \hat{q}  ( {\bf k} )   \right] } ,
\end{eqnarray}
where we have assumed that $\mathcal{L}$  does not cross  the nodal lines
and ${\bf K}_1$ and ${\bf K}_2$ denote the two
time-reversal-invariant
momenta on the path $\mathcal{L}$. Repeating similar steps as in the
previous subsection, we find that for a weak-pairing superconductor
$N^{\textrm{1D}}_{\mathcal{L}}$ simplifies to
\begin{eqnarray} \label{1dZ2second}
N^{\textrm{1D}}_{\mathcal{L}}
=
\sgn \big( \Delta^+_{\mathcal{L}} \big) \sgn \big( \Delta^-_{\mathcal{L}} \big) ,
\end{eqnarray}
where $\sgn ( \Delta^\pm_{\mathcal{L}} )$ represents the sign of the gap at
the points given by the intersection of $\mathcal{L}$ with the
positive/negative-helicity Fermi surface.

\subsection{Topological criteria for the existence\\of zero-energy surface states}
\label{sec:topCriteria}

As a consequence of a bulk-boundary
correspondence,\cite{schnyder2008,ryuHatsugai} a nontrivial value of any of
the three topological invariants 
\eqref{eq: 1D winding no}, \eqref{Z2noLattice}, and \eqref{1dZ2first} signals
the appearance of zero-energy states at the surface 
of the NCS. That is, $W_{\mathcal{L}} \ne  0$ leads to surface flat bands,
$\tilde{N}^{\textrm{2D}}_{\pm} = -1$ 
gives rise to arc surface states, and $N^{\textrm{1D}}_{\mathcal{L}} = -1$
results in Majorana modes at time-reversal-invariant momenta of the surface
BZ. 
In the following, we discuss in detail the topological criteria for the
existence of these three types of surface states. For that purpose,
we denote the coordinates parallel (perpendicular) to a given surface of the
NCS by ${\bf r}_{\parallel}$ ($r_{\perp}$) and the corresponding momenta by
${\bf k}_{\parallel}$~($k_{\perp}$).  

\paragraph{Surface flat bands:} 

The appearance of surface flat bands can be understood by considering a  
continuous deformation of the closed integration path $\mathcal{L}$ of
$W_{\mathcal{L}}$, Eq.~\eqref{eq: 1D winding no}, 
into an infinite semicircle, such that the diameter of the semicircle contains
the line $(k_{\perp}, {\bf k}_{\parallel}$), with $k_{\perp} \in \mathbbm{R}$
and  ${\bf k}_{\parallel}$ fixed.  
This path deformation does not alter the value of the integral, as long as no
nodal line is crossed while deforming $\mathcal{L}$; if a nodal line is
crossed $W_{\mathcal{L}}$ changes by $\pm 1$. As can be seen from
Eq.~\eqref{WLsimpI}, the integral along the arc of the semicircle is zero and
hence the integral along the diameter is given by\cite{approximationsI}   
\begin{equation} \label{topCritFlatBand}
W_{(lmn)} ( {{\bf k}_{\parallel} } )
= - \frac{1}{2}
\sum_{\nu = \pm }
\Big[
\sgn \big( \Delta^{\nu}_{{\bf k}_{F,\nu}} \big)
  - \sgn \big( \Delta^{\nu}_{\widetilde{\bf k}_{F,\nu}} \big)
\Big] , 
\end{equation}
where ${\bf k}_{F,\pm} = (k_{\perp,\pm},{\bf k}_\parallel)$ and $\widetilde{\bf
  k}_{F,\pm} = (\widetilde{k}_{\perp,\pm},{\bf k}_\parallel)$ satisfy  
$\xi^{\pm}_{{\bf k}_{F,\pm}}=0$ and $\xi^{\pm}_{\widetilde{\bf k}_{F,\pm}}=0$,
respectively, with ${\bf k}_{F,\pm}$ ($\widetilde{\bf  k}_{F,\pm}$)
corresponding to solutions with positive (negative) 
signs of the Fermi-velocity component perpendicular to the surface.
The subscript $(lmn)$ in Eq.~\eqref{topCritFlatBand} parametrizes the
direction perpendicular to the surface. 
From Eq.~\eqref{topCritFlatBand} it follows that zero-energy surface states
occur whenever $W_{(lmn)} ( {{\bf k}_{\parallel}} ) \ne 0$, which
corresponds to regions of the surface BZ that are bounded by the projections
of the nodal rings of the bulk gap.

\paragraph{Arc surface states:}

Consider an NCS satisfying symmetry~\eqref{newTRS} for $E_{k} ( \hat{\bf u})$,
with $k \in \mathbbm{R}$, and with two nodal rings 
carrying nontrivial $\mathbbm{Z}_2$ topological charge,
Eq.~\eqref{Z2ChArge}.  Any two-dimensional subsystem $E_{k} ( \hat{\bf u})$ 
lying between these two line nodes can be viewed as a
time-reversal-invariant topological superconductor in class DIII. When 
these two-dimensional subsystems are terminated by a boundary, there appear
helical Majorana edge states. These Majorana modes cross zero energy on the
surface BZ, somewhere in between the projections of the two nodal rings. Thus,
there is a one-dimensional arc of zero-energy states connecting the
projections of the two nodal rings.  
If the projected nodal rings have a finite overlap in the surface BZ,
cancellation occurs and no arc surface states are expected.  
By symmetry, the arc surface states always appear at those
surface momenta that are invariant under symmetry~\eqref{newTRS}. 
It is interesting to note that these arc surface states are reminiscent
  of the Fermi arcs that 
appear at the surface of Weyl semimetals, which have recently been discussed
in the  context of the A phase of ${}^3$He,\cite{volovikHe3A} pyrochlore
iridates,\cite{wanWeyl2011} and topological insulator
multilayers.\cite{burkovBalents2011}

\paragraph{Majorana surface states:} 

We observe that the gap~$\Delta^{\nu}_{\bf k}$ (with $\nu = \pm$) has the
same values at any two momenta related 
by TRS. Thus, for a given time-reversal-invariant  surface
momentum ${\bf K}_{\parallel}$, 
 the one-dimensional $\mathbbm{Z}_2$ number~\eqref{1dZ2second}
 reads\cite{approximationsI} 
\begin{eqnarray} \label{topCritMajo}
N^{\textrm{1D}}_{{\bf K}_{\parallel}}
=
\sgn \big( \Delta^+_{{\bf K}_{F,+}} \big) \sgn \big( \Delta^-_{{\bf K}_{F,-} } \big) ,
\end{eqnarray}
where ${\bf K}_{F,\pm} = ( K_{{\perp}\pm}, {\bf K}_{\parallel} )$ 
satisfies $\xi^{\pm}_{{\bf K}_{F,\pm}} = 0$ and the sign
of the Fermi-velocity component $\left. \partial_{K_\perp} \xi^{\pm}_{\bf k}
\right|_{{\bf k}= {\bf K}_{F,\pm} }$ is assumed to be positive. 
Consequently, $N^{\textrm{1D}}_{{\bf K}_{\parallel}}=-1$ indicates the
presence of Kramers-degenerate Majorana surface modes at the
time-reversal-invariant momentum~${\bf K}_{\parallel}$ of the surface BZ.

Before closing this subsection, we note that the topological protection of
boundary modes in nodal NCSs is weaker
than the protection of surface states in strong topological insulators or
superconductors with a full gap in the bulk. Nevertheless, the surface states
in nodal NCSs are expected to possess a certain
robustness against disorder, similar to the zero-energy edge states in
graphene\cite{wimmerPRB10} or in d$_{x^2-y^2}$-wave
superconductors.\cite{matsumotoJPSJ95} For example, we expect that weak
  surface 
  roughness will result in a broadening of the surface states described
  above, but the surface spectral function will still display sharp
  peaks at the bound state energies. Deeper examination of the effect of
  disorder is beyond the scope of this paper, and is left to later
  publications.

\section{Bound state spectra}
\label{sec:boundStates}

In this section, we use quasiclassical scattering theory to derive 
conditions for the existence of surface bound states. We will see that the
quasiclassical criteria 
for the appearance of zero-energy surface states are in perfect agreement with
the topological criteria given in the previous section.  
Hence, this provides a verification of the bulk-boundary correspondence
that relates   
nontrivial topological characteristics of quasiparticle wavefunctions in the
bulk to the existence of zero-energy states at the surface.

\subsection{General condition for bound states}

The wavevector component ${\bf k}_\parallel$ parallel to an ideal surface is 
a good quantum number due to translational invariance, and so we can construct 
the bound-state wavefunction for each point in the surface BZ
independently. For a given ${\bf k}_\parallel$, the usual quasiclassical
method~\cite{kashiwaya_tanaka00} is only applicable if there are at most two
solutions in each helicity sector.
If there are more than two solutions in either sector
(for example for a concave Fermi surface) we are unable to uniquely determine the
coefficients of the different spinors appearing in the
bound-state-wavefunction ansatz for the given ${\bf k}_\parallel$. Our approach is
therefore reasonable in the 
limit of weak to moderate SOC, where the concavity of
the Fermi surface is only relevant over a small fraction
of the surface BZ.

For given ${\bf k}_\parallel$ we must therefore in general consider four
wavevectors ${\bf k}_\pm = (k_{\perp,\pm},{\bf k}_\parallel)$ and $\widetilde{\bf
  k}_\pm = (\widetilde{k}_{\perp,\pm},{\bf k}_\parallel)$, which satisfy
$\xi^{\pm}_{{\bf k}_\pm}=\xi^{\pm}_{\widetilde{\bf k}_\pm}=0$. We classify a
solution as propagating if $k_\perp$ is real and as evanescent otherwise.
In the former case, ${\bf k}_\pm$ and $\widetilde{\bf
  k}_\pm$ correspond to solutions with opposite signs of the Fermi-velocity
component perpendicular to the surface. We assume that the NSC is located in
the half-space  $r_\perp>0$.

\subsubsection{Propagating solutions on both Fermi surfaces} \label{sec:2prop}

If there are propagating solutions on both the positive-helicity and
the negative-helicity Fermi surfaces we have the wavefunction ansatz
\begin{subequations}\label{eq:Psicase1}
\beqarray
\Psi({\bf k}_\parallel; {\bf r}) & = & \sum_{\nu = \pm}\Psi_{\nu}({\bf
  k}_\parallel;{\bf r})e^{i{\bf
    k}_\parallel\cdot{\bf r}}\,, \\
\Psi_{\nu}({\bf
  k}_\parallel;{\bf r}) & = & \sum_{{\bf k}={\bf k}_\nu,\widetilde{\bf
    k}_\nu}\alpha_{\nu}({\bf k})\psi_{\nu}( {\bf
  k})e^{ik_{\perp} r_\perp}e^{-\kappa^{\nu}_{{\bf k}} r_\perp}\,, 
\eeqarray
where the positive-helicity and negative-helicity spinors are given by
\beqarray
\psi_{+}({\bf k}) &=& \left(\begin{array}{cccc}1, & \frac{l^{x}_{\bf k} +
    il^{y}_{\bf k}}{|{\bf l}_{\bf k}| + l^{z}_{\bf k}}, & -\frac{l^{x}_{\bf k}
    + il^{y}_{\bf k}}{|{\bf l}_{\bf k}| + l^{z}_{\bf k}}\, \gamma^{+}_{\bf k}, &
  \gamma^{+}_{\bf k}\end{array}\right)^{T} \,,
\\
\psi_{-}({\bf k}) &=& \left(\begin{array}{cccc}\frac{l^{x}_{\bf k} -
    il^{y}_{\bf k}}{|{\bf l}_{\bf k}| + l^{z}_{\bf k}}, & -1, & \gamma^{-}_{\bf k}, &
  \frac{l^{x}_{\bf k}
    - il^{y}_{\bf k}}{|{\bf l}_{\bf k}| + l^{z}_{\bf k}}\, \gamma^{-}_{\bf
    k}\end{array}\right)^{T}\, ,
\eeqarray
respectively, with 
\begin{eqnarray}
\gamma^{\pm}_{\bf k} 
&=& 
\frac{1}{\Delta^\pm_{\bf k}}\left[E -
  i \sgn(v^{\pm}_{F,\perp}({\bf k}))\sqrt{|\Delta^\pm_{\bf k}|^2 - E^2}\right]\,, \\
\kappa^{\pm}_{\bf k}
&=&
\frac{1}{\hbar|v^{\pm}_{F,\perp}({\bf k})|}\sqrt{|\Delta^{\pm}_{\bf k}|^2-E^2}\,,
\end{eqnarray}
\end{subequations}
and $v^{\nu}_{F,\perp}({\bf k})$ is the component of the $\nu=\pm$ helicity
Fermi velocity normal to the surface. 

A bound state is realized when it is possible to choose the coefficients
$\alpha_{\nu}({\bf k})$ in~\eq{eq:Psicase1} such that the wavefunction vanishes
at the surface, i.e., $\Psi({\bf k}_\parallel; {\bf
  r})|_{r_\perp=0} = 0$. This is equivalent to the condition
\begin{widetext}
\beqarray
0 & = & \big(\gamma^{-}_{{\bf k}_{-}} - \gamma^{-}_{\widetilde{{\bf
      k}}_{-}}\big)\big(\gamma^{+}_{\widetilde{\bf k}_{+}} - \gamma^{+}_{{\bf k}_{+}}\big)\left[\big(|{\bf l}_{{\bf k}_+}| + l^{z}_{{\bf k}_{+}}\big)\big(|{\bf l}_{\widetilde{\bf k}_-}| +
  l^{z}_{\widetilde{\bf k}_{-}}\big) - \big(l^{x}_{{\bf k}_{+}} + il^{y}_{{\bf k}_{+}}\big)\big(-l^{x}_{\widetilde{\bf k}_{-}} + il^{y}_{\widetilde{\bf
      k}_{-}}\big)\right] \notag \\
&& \times\left[\big(|{\bf
    l}_{{\bf k}_-}| + l^{z}_{{\bf k}_{-}}\big)\big(|{\bf l}_{\widetilde{\bf k}_+}| +
  l^{z}_{\widetilde{\bf k}_{+}}\big) - \big(-l^{x}_{{\bf k}_{-}} +
  il^{y}_{{\bf k}_{-}}\big)\big(l^{x}_{\widetilde{\bf k}_{+}} + il^{y}_{\widetilde{\bf
      k}_{+}}\big)\right] \notag \\
&& + \big(\gamma^{-}_{{\bf k}_{-}} -
\gamma^{+}_{\widetilde{{\bf k}}_{-}}\big)\big(\gamma^{+}_{{\bf k}_{+}} -
\gamma^{-}_{\widetilde{\bf k}_{-}}\big)\left[\big(|{\bf l}_{{\bf k}_+}| + l^{z}_{{\bf
      k}_{+}}\big)\big(l^{x}_{\widetilde{\bf k}_{+}} + il^{y}_{\widetilde{\bf 
      k}_{+}}\big) - \big(l^{x}_{{\bf k}_{+}} + il^{y}_{{\bf k}_{+}}\big)\big(|{\bf
    l}_{\widetilde{\bf k}_+}| + l^{z}_{\widetilde{\bf k}_{+}}\big)\right] \notag
\\
&& \times \left[\big(|{\bf l}_{{\bf k}_-}| + l^{z}_{{\bf
      k}_{-}}\big)\big(-l^{x}_{\widetilde{\bf k}_{-}} + il^{y}_{\widetilde{\bf
      k}_{-}}\big) - \big(-l^{x}_{{\bf k}_{-}} +
  il^{y}_{{\bf k}_{-}}\big)\big(|{\bf l}_{\widetilde{\bf k}_-}| +
  l^{z}_{\widetilde{\bf k}_{-}}\big)\right]. \label{eq:cond2prop}
\eeqarray 
\end{widetext}
Solutions of this equation satisfying
$|E|<\min\{|\Delta^{\pm}_{{\bf
    k}_\pm}|,|\Delta^{\pm}_{\widetilde{\bf k}_\pm}|\}$ are
the bound-state energies. Focusing on 
zero-energy solutions that occur within a finite region of the surface
BZ, we find three possibilities for such states:
\begin{eqnarray}
\textrm{(i)} \; && \sgn (\Delta^{-}_{{\bf k}_-} \Delta^{-}_{\widetilde{\bf k}_-} ) = -1 
\;  \textrm{and}  \; \sgn ( \Delta^{+}_{{\bf k}_+} \Delta^{+}_{\widetilde{\bf k}_+} ) = +1,
  \nonumber\\
\textrm{(ii)} \; &&  \sgn (\Delta^{-}_{{\bf k}_-} \Delta^{-}_{\widetilde{\bf k}_-} ) = +1 
 \;  \textrm{and} \; \sgn (\Delta^{+}_{{\bf k}_+} \Delta^{+}_{\widetilde{\bf k}_+}) =-1,
   \nonumber\\
\textrm{(iii)} \; && \sgn ( \Delta^{-}_{{\bf k}_-} \Delta^{-}_{\widetilde{\bf k}_-} ) = -1
\;   \textrm{and} \; \sgn (\Delta^{+}_{{\bf k}_+} \Delta^{+}_{\widetilde{\bf k}_+} ) = -1 
\;   \textrm{and}
   \nonumber\\  
   && \sgn( \Delta^{-}_{{\bf k}_-} \Delta^{+}_{\widetilde{\bf k}_+} ) = -1 .
  \nonumber
\end{eqnarray}
These conditions agree perfectly with the topological criterion
\eqref{topCritFlatBand}. That is,
scenarios~(i) and~(ii) correspond to topologically protected singly degenerate
zero-energy 
states with winding number $W_{(lmn)}=\pm1$, while (iii) gives
doubly degenerate states of winding number $W_{(lmn)} = \pm2$. We
note that a careful examination of the terms 
involving ${\bf l}_{\bf k}$ can also yield zero-energy \emph{dispersing} 
states at isolated points or in a line. These states include, but are not
limited to, the topologically protected Majorana modes and arc surface
states introduced in~\Sec{sec:topCriteria}. The general form of this condition is
rather cumbersome, but it simplifies significantly in the 
limit of weak SOC and so we defer discussion to this case.

\subsubsection{Propagating solutions on only one Fermi surface}

In the case that there are propagating solutions only on the
negative-helicity Fermi surface, the negative-helicity components of the wavefunction
ansatz remain as above, but the positive-helicity components are now written
\begin{subequations}
\beqarray
\Psi_+({\bf k}_\parallel; {\bf r}) &=&
 \Big\{\alpha_{+}({\bf p})\phi({\bf
  p})e^{-iq_{{\bf p}} r_\perp}  \notag \\
&& {} + \widetilde{\alpha}_{+}({\bf p})\widetilde{\phi}({\bf
  p})e^{iq_{{\bf p}} r_\perp} \Big\}\, e^{ip_{\perp} r_\perp} \, , \label{eq:Psicase2}
\eeqarray
where ${\bf p}$ is chosen from ${\bf k}_+$ and $\widetilde{\bf
  k}_+$ such that the imaginary part of $p_\perp$ is positive, the spinors
$\phi({\bf p})$ and $\widetilde{\phi}({\bf p})$ are defined by
\beqarray
\phi({\bf p}) & = & \left(\begin{array}{cccc}1,
  \frac{l^{x}_{\bf p}+il^{y}_{\bf p}}{|{\bf l}_{\bf p}| + l^{z}_{\bf p}}, &
  -\frac{l^{x}_{\bf p}+il^{y}_{\bf p}}{|{\bf l}_{\bf p}| + l^{z}_{\bf
      p}}\Gamma_{\bf p}, & \Gamma_{\bf
    p}\end{array}\right)^{T}\,, \\
\widetilde{\phi}({\bf p}) & = & \left(\begin{array}{cccc}1,
  \frac{l^{x}_{\bf p}+il^{y}_{\bf p}}{|{\bf l}_{\bf p}| + l^{z}_{\bf p}}, &
  -\frac{l^{x}_{\bf p}+il^{y}_{\bf p}}{|{\bf l}_{\bf p}| + l^{z}_{\bf
      p}}\widetilde{\Gamma}_{\bf p}, & \widetilde{\Gamma}_{\bf
    p}\end{array}\right)^{T}\,,
\eeqarray
and we have
\beqarray
\Gamma_{\bf p} &=& \frac{1}{\Delta^{+}_{\bf p}}\left(E + i\sqrt{(\Delta^{+}_{\bf
      p})^2-E^2}\right)\,, \\
\widetilde{\Gamma}_{\bf p} &=& \frac{1}{\Delta^{+}_{\bf p}}\left(E  -
i\sqrt{(\Delta^{+}_{\bf p})^2-E^2}\right) \,, \\
q_{\bf p} &=& \frac{1}{\hbar\,{\rm Im}\{v^{+}_{F,\perp}({\bf
    p})\}}\sqrt{(\Delta^{+}_{\bf p})^2 - E^2}\,. 
\eeqarray
\end{subequations}
Using the same argument as above, we find the condition
\begin{widetext}
\beqarray
0 & = & \big(\widetilde{\Gamma}_{{\bf p}} - \Gamma_{{\bf p}}\big)\big(\gamma^{-}_{{\bf
    k}_{-}} - \gamma^{-}_{\widetilde{\bf k}_{-}}\big)\left[\big(|{\bf l}_{{\bf p}}| +
  l^{z}_{{\bf p}}\big)\big(|{\bf 
    l}_{{\bf k}_-}| + l^{z}_{{\bf k}_{-}}\big)-\big(l^{x}_{{\bf p}} +
  il^{y}_{{\bf p}}\big)\big(-l^{x}_{{\bf k}_{-}} + il^{y}_{{\bf k}_{-}}\big)\right] \notag
\\
&& \times\left[\big(|{\bf l}_{{\bf p}}| +
  l^{z}_{{\bf p}}\big)\big(|{\bf l}_{\widetilde{\bf k}_-}| +  
  l^{z}_{\widetilde{\bf k}_{-}}\big) - \big(l^{x}_{{\bf p}} +
  il^{y}_{{\bf p}}\big)\big(-l^{x}_{\widetilde{\bf k}_{-}} + il^{y}_{\widetilde{\bf
      k}_{-}}\big)\right] \label{eq:cond1prop}
\eeqarray
\end{widetext}
for the formation of bound states.
Unlike~\eq{eq:cond2prop}, this condition only allows for the existence of
\emph{singly} degenerate  zero-energy surface states which occur whenever
$\sgn (\Delta^{-}_{{\bf k}_-}) = 
-\sgn (\Delta^{-}_{\widetilde{\bf k}_-})$.\cite{footnoteTRIM} 
Again, we find that this criterion matches with the topological
one, Eq.~\eqref{topCritFlatBand}.

\subsubsection{Limit of weak spin-orbit coupling} \label{sec:weakSOC}

Because of the difficulty of working with spin-orbit split Fermi surfaces, many
theoretical studies assume that the SOC is weak so that
the spin splitting of the Fermi surfaces can be
ignored.~\cite{Iniotakis2007,Vorontsov2008,eschrigIniotakisReview,Vorontsov2008b,Brydon2011,Lu2010}  
This limit can be directly obtained from~\eq{eq:cond2prop} by setting ${\bf
  k}_+ = {\bf k}_- = {\bf k}$ and $\widetilde{\bf k}_+=\widetilde{\bf
  k}_-=\widetilde{\bf k}$ 
to yield the compact bound-state condition
\beqarray \label{eq:conddegen}
0 & = & \big(\gamma^{+}_{\widetilde{\bf k}} - \gamma^{-}_{{\bf
      k}}\big)\big(\gamma^{-}_{\widetilde{\bf k}} - \gamma^{+}_{{\bf k}}\big)\big(|{\bf
  l}_{\bf k}||{\bf l}_{\widetilde{\bf k}}|-{\bf l}_{\bf k}\cdot {\bf
  l}_{\widetilde{\bf k}}\big)  
   \notag \\
&& + \big(\gamma^{-}_{\widetilde{\bf k}} - \gamma^{-}_{{\bf
      k}}\big)\big(\gamma^{+}_{\widetilde{\bf k}} - \gamma^{+}_{{\bf k}}\big)\big(|{\bf
  l}_{\bf k}||{\bf l}_{\widetilde{\bf k}}|+{\bf l}_{\bf k}\cdot   {\bf
  l}_{\widetilde{\bf k}}\big) . 
\eeqarray
This result was previously presented in~\Ref{Brydon2011} and is utilized here
to obtain the surface bound-state spectra for the  $O$ and $T_{d}$ point 
groups.  Although describing a physically idealized situation,
Eq.~\eqref{eq:conddegen} is useful as it significantly simplifies the
discussion of the zero-energy \emph{dispersing} states. In particular, we find
that such states are possible if 
(i)  
${\bf l}_{\bf k}\cdot {\bf  l}_{\widetilde{\bf k}} = | {\bf l}_{\bf k} | | {\bf l}_{\widetilde{\bf k}} |$
  and 
  $\sgn (\Delta^{+}_{\bf k}) = \sgn (\Delta^{-}_{\widetilde{\bf k}})
  \neq \sgn (\Delta^{+}_{\widetilde{\bf k}}) = \sgn (\Delta^{-}_{\bf
    k})$ 
    or (ii) 
${\bf l}_{\bf k}\cdot {\bf  l}_{\widetilde{\bf k}} = - | {\bf l}_{\bf k} | | {\bf l}_{\widetilde{\bf k}} |$
  and
  $\sgn (\Delta^{+}_{\bf k}) = \sgn (\Delta^{+}_{\widetilde{\bf k}})
  \neq \sgn (\Delta^{-}_{\bf k}) = \sgn (\Delta^{-}_{\widetilde{\bf
    k}})$.
Scenario (ii) includes, but is not limited to, the topological
criteria \eqref{Z2simp} and
\eqref{topCritMajo} which desrcibe arc surface states and the
Kramers-degenerate Majorana modes, respectively. In particular, the
antisymmetry of the SOC vector ensures that the latter state is realized at 
the zone center. In contrast, states satisfying scenario (i) do not fall
into either topological category: They cannot be Kramers-degenerate Majorana modes,
since the requirement that ${\bf l}_{\bf k}\cdot {\bf  l}_{\widetilde{\bf k}}
= | {\bf l}_{\bf k} | | {\bf l}_{\widetilde{\bf k}} |$ is never satisfied at
the zone center; nor can they be arc surface states protected by a
$\mathbb{Z}_2$ number, as the condition on the gap signs implies that any
plane containing ${\bf k}$ and $\widetilde{\bf k}$ must also
contain a gap node. We note that the condition on the gaps is equivalent to
$\sgn(f({\bf k}))=-\sgn(f(\widetilde{\bf k}))$, which is not realized together
with the condition on the polarization vector for any of the systems 
considered in this paper, i.e., all examples of zero-energy dispersing
states presented below satisfy scenario (ii).

\begin{figure}[t!]
  \includegraphics[clip,width=0.75\columnwidth]{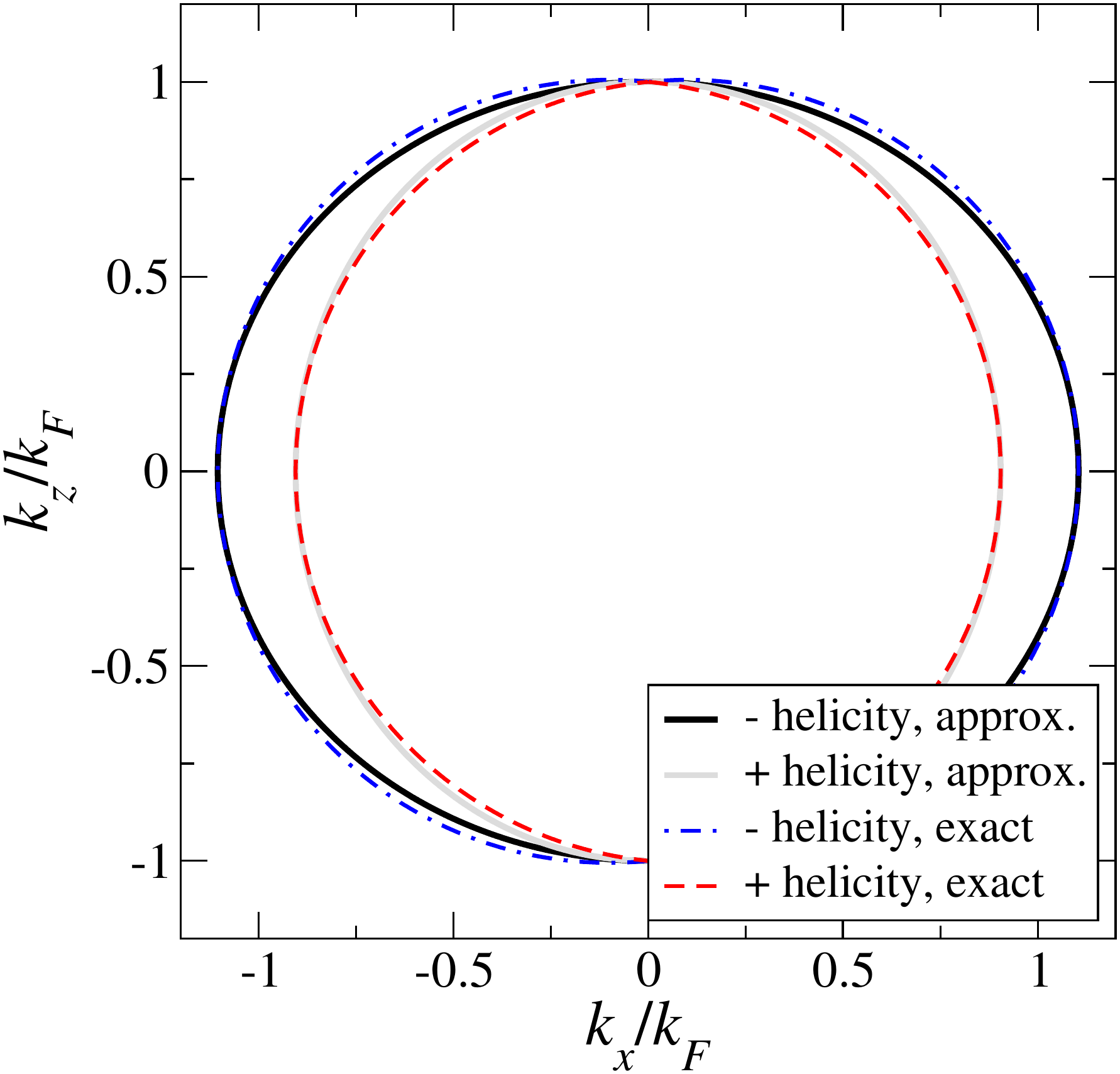}
  \caption{\label{FS}  (Color online) Comparison of the Fermi surfaces in
      the \mbox{$x$-$z$} plane of
      the approximate dispersion~\eq{eq:approxdisp}
      (solid lines) and the exact dispersion~\eq{eq:exactdisp} (dashed lines)
      for $\lambda{k_F}/2\mu=0.1$.
}
\end{figure}

\subsection{Surface states for the tetragonal point group  $C_{4v}$}

In this section we analyze the bound states of the $C_{4v}$ point group for
finite spin-orbit splitting of the Fermi surface. Assuming a spherical
Fermi surface in the limit of vanishing SOC
we find the normal-state dispersion  at finite $\lambda$,
\beq
\xi^{\pm}_{\bf k} = \frac{\mu}{k_F^2}|{\bf k}|^2 - \mu \pm
\lambda\sqrt{k_x^2 + k_y^2}\,, \label{eq:exactdisp}
\eeq
where $k_{F}= \sqrt{ 2m\mu/\hbar^2 }$. This implies a concave negative-helicity Fermi 
surface near $k_z=\pm|k_F|$, and so our quasiclassical method may not be 
well-defined across the entire surface BZ. In order to avoid this
problem, we approximate the negative-helicity and positive-helicity Fermi
surfaces by oblate and prolate spheroids, respectively,
\beq
\xi^{\pm}_{\bf k} \approx  \frac{\mu}{(k_{F}^\pm)^2}(k_x^2 + k_y^2) +
\frac{\mu}{k_{F}^2}k_z^2 -\mu \label{eq:approxdisp} ,
\eeq
where  
\beq
k^{\pm}_{F} = k_{F}\left[\mp\frac{\lambda k_F}{2\mu} + \sqrt{1 +
    \left(\frac{\lambda k_F}{2\mu}\right)^2}\right] 
\eeq
gives the radius of the helical Fermi surfaces in the $k_z=0$
plane. Our dispersion, \eq{eq:approxdisp}, qualitatively captures the
salient features of the true dispersion $\xi^{\pm}_{\bf
  k}$, \eq{eq:exactdisp}: In the $k_z=0$ plane the Fermi surfaces are circles
of different radii, and they only touch along the line  $k_x=k_y=0$. As can
be seen in~\fig{FS}, the Fermi surfaces of the approximate and exact
dispersions agree very well for moderate $\lambda{k_F}/2\mu\leq0.1$. To
determine the bound-state spectra at given ${\bf k}_\parallel$, we need only
find ${\bf k}_{\nu}$, $\widetilde{\bf k}_{\nu}$, and the sign of
$v^{\nu}_{F,\perp}({\bf k})$ at these points so that we expect that obtaining
these parameters from~\eq{eq:approxdisp} should give excellent agreement with
the exact results across most of the surface BZ.

\begin{figure}
  \includegraphics[clip,width=\columnwidth]{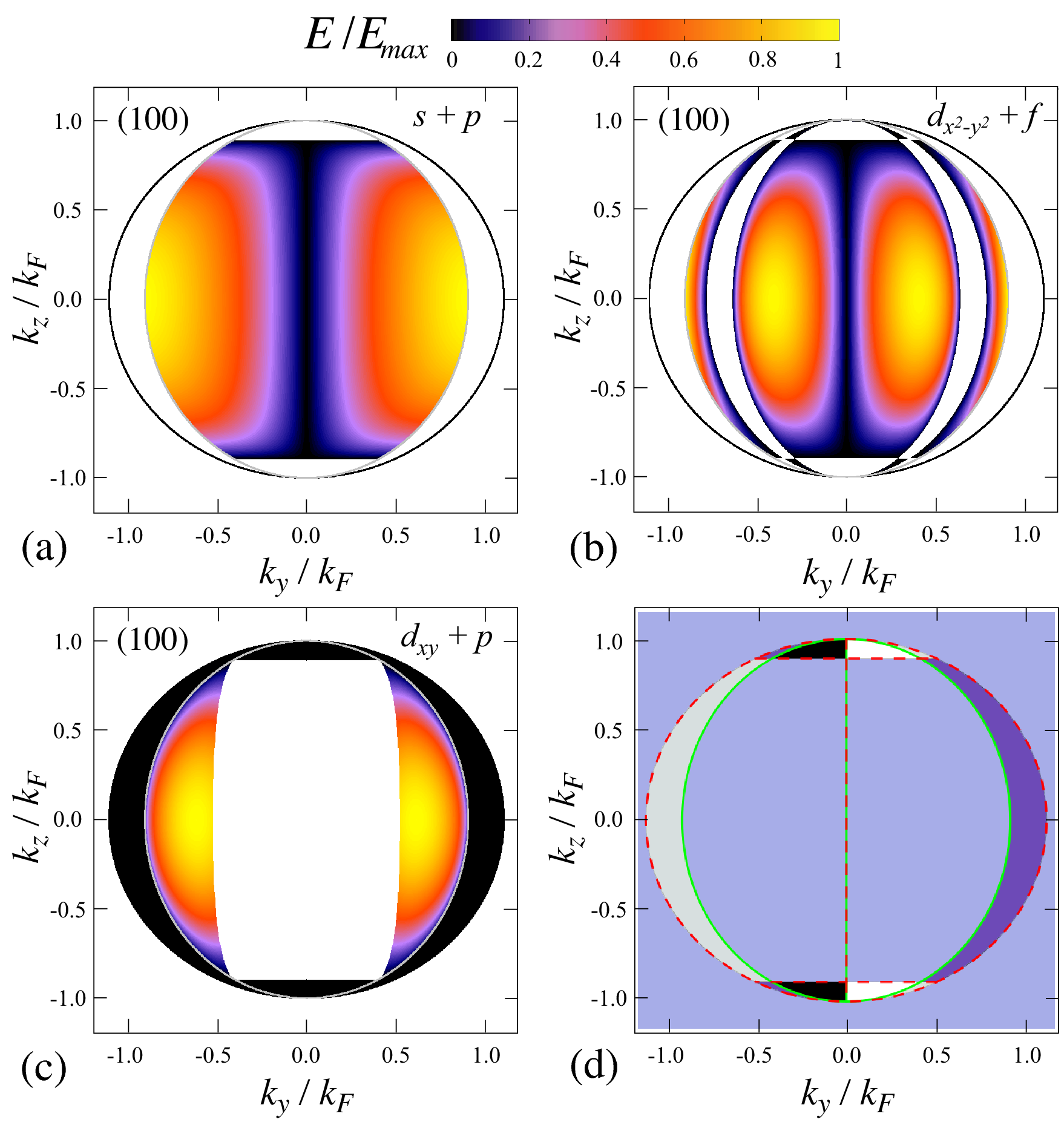}
  \caption{\label{fig1} (Color online) Surface bound-state spectra
    at the (100) face of a $C_{4v}$ point-group NCS as a function of
    surface momentum ${\bf k}_{\parallel} = (k_y, k_z)$ with (a) $(s+p)$-wave,
    (b) $(d_{x^2-y^2} + f)$-wave, and (c) $(d_{xy} + p)$-wave pairing
    symmetry. Here we set $\lambda k_F / (2 \mu ) = 0.1$ and $r=0.5$. The
    color scale 
    indicates the energy: black represents zero energy while yellow
    represents the maximum energy $E_{\textrm{max}}$.
    The black (gray) line shows the extent of
    the projected negative-helicity (positive-helicity) Fermi surface. (d)
    Winding number $W_{\textrm{(100)}}$, Eq.~\eqref{topCritFlatBand}, at the
    (100) face corresponding to the
    same parameters as in panel (c). Black (white) indicates
$W_{\textrm{(100)}}= + 2$ ($-2$), dark blue (gray) corresponds to
$W_{\textrm{(100)}} = + 1$ ($-1)$, while light blue is $W_{\textrm{(100)}}
= 0$. The red dashed (green solid) lines represent the nodal lines on the
negative-helicity (positive-helicity) Fermi surface.  
}
\end{figure}

\begin{figure}
\includegraphics[clip,width=\columnwidth]{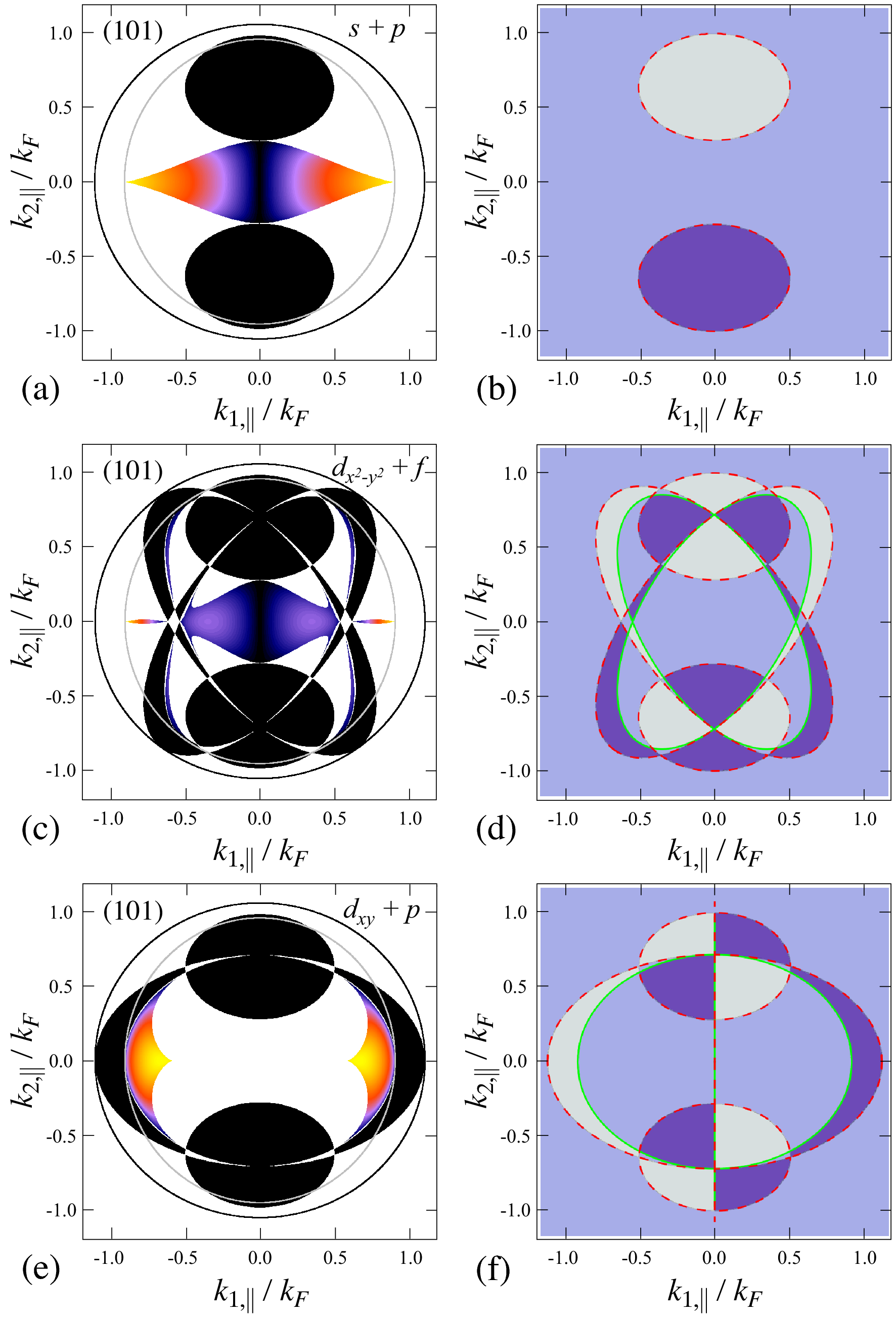}
  \caption{\label{fig2}
  (Color online) Surface bound-state spectra at the (101) face of a $C_{4v}$ point-group NCS
as a function of surface momentum ${\bf k}_{\parallel}$ with (a) $(s+p)$-wave,  
(c) $(d_{x^2-y^2} + f)$-wave, and (e) $(d_{xy} + p)$-wave pairing symmetry.
Here we set $\lambda k_F / (2 \mu) = 0.1$ and $r=0.5$. The color scale is the same as in
Figs.~\ref{fig1}(a)--(c). The black (gray) line shows the extent of the projected
negative-helicity (positive-helicity) Fermi surface. Panels (b), (d), and (f)
show the winding
number $W_{\textrm{(101)}}$ at the (101) face corresponding to the same
parameters as in panels (a), (c), and (e), respectively. Dark blue (gray)
indicates $W_{\textrm{(101)}} = + 1$ ($-1)$, while light blue is
$W_{\textrm{(101)}} = 0$. The red dashed (green solid) lines represent the
nodal lines on the  negative-helicity (positive-helicity) Fermi surface.  
}  
\end{figure}

In~\figs{fig1}(a)--(c) we present the surface bound-state spectra
for the 
(100) face of a $C_{4v}$ point-group NCS for several different pairing
symmetries. In the case of $(s+p)$-wave pairing we observe dispersing
  states
  in the region bounded by the projected positive-helicity Fermi surface and
  the two nodal rings at $k_z \simeq \pm 0.89 k_F$  [\fig{fig1}(a)]. Since
  these nodal rings  
carry nontrivial $\mathbbm{Z}_2$ topological charge, in agreement with the
discussion in Sec.~\ref{sec:topCriteria} there are zero-energy arc surface
states~\cite{Tanaka2009,Vorontsov2008,Vorontsov2008b,Lu2010,Iniotakis2007,eschrigIniotakisReview,Brydon2011} 
connecting the projections of the two nodal rings.
Moreover, in accordance
with the analysis of Secs.~\ref{sec:topCriteria} and~\ref{sec:weakSOC}, 
we find that this arc lies along the line 
$k_y=0$, i.e., at surface momenta that are left invariant
under symmetry \eqref{newTRS}, and where we have
${\bf l}_{\bf k}\cdot {\bf  l}_{\widetilde{\bf k}} = - | {\bf l}_{\bf k} | | {\bf l}_{\widetilde{\bf k}} |$.

The situation is qualitatively similar for the $(d_{x^2-y^2}+f)$-wave case
[\fig{fig1}(b)], although the extra nodes due to the $d_{x^2-y^2}$ form-factor
$f({\bf k})$ modulate the results of the $(s+p)$-wave case. Note
that these extra nodes remove the condition for the topological protection of
the line of zero-energy surface states at $k_y=0$; the zero energy state
at the zone center nevertheless remains a Kramers-degenerate Majorana mode.
In contrast, \fig{fig1}(c) shows that we do not find any dispersing
zero-energy states for the $(d_{xy}+p)$-wave pairing, but instead there
are zero-energy flat bands in 
several regions bounded by the projected line nodes of the
positive-helicity and negative-helicity gaps. The zero-energy states lying outside the
projected positive-helicity Fermi surface (light gray line) are the
singly degenerate 
``time-reversal-invariant Majorana states'' found in~\Ref{Tanaka2010}, and are
associated with a non-trivial winding number of $W_{\textrm{(100)}}=\pm1$ as shown
in~\fig{fig1}(d). In contrast, 
the zero-energy surface states lying inside the projected positive-helicity
Fermi surface are doubly degenerate and have winding number
$W_{\textrm{(100)}}=\pm2$. These states occur in the region where the gap has
predominantly singlet character, and hence are due to the same mechanism as the
zero-energy surface states in a 
pure $d_{xy}$-wave superconductor.~\cite{hu1994,ryuHatsugai,kashiwaya_tanaka00}

The bound states at the (101) surface shown in~\fig{fig2} display a much
more interesting topological character. We first consider the results for the
$(s+p)$-wave case [\fig{fig2}(a)], which are qualitatively identical to those
previously obtained in~\Ref{Brydon2011} for vanishing spin-orbit splitting of
the Fermi surfaces. Specifically, we find that flat 
zero-energy bands occur within the projected nodes of the negative-helicity 
Fermi surface where the sign of the
negative-helicity gap reverses between the forward- and backward-facing halves
of the Fermi surface. 
These zero-energy states are associated with a finite winding number
$W_{(101)}=\pm1$; the 
variation of $W_{(101)}$ across the surface BZ shown in
panel (b) clearly demonstrates the non-zero topological 
charge associated with the nodal rings of the negative-helicity gap. Like for
the (100) face, the projections of these topologically charged nodal rings
are connected by arc surface states.
The presence of higher angular-momentum harmonics
[\figs{fig2}(c)--(f)] results 
in the appearance of additional regions of zero-energy states due to the
nodes of 
both the positive-helicity and negative-helicity gaps. All of these states correspond
to a winding number of $W_{(101)}=\pm1$, 
as can be seen by comparing the bound state spectra,~\fig{fig2}(c) and (e),
with the winding number calculations,~\fig{fig2}(d) and (f), respectively. 

\begin{figure}
  \includegraphics[clip,width=0.9\columnwidth]{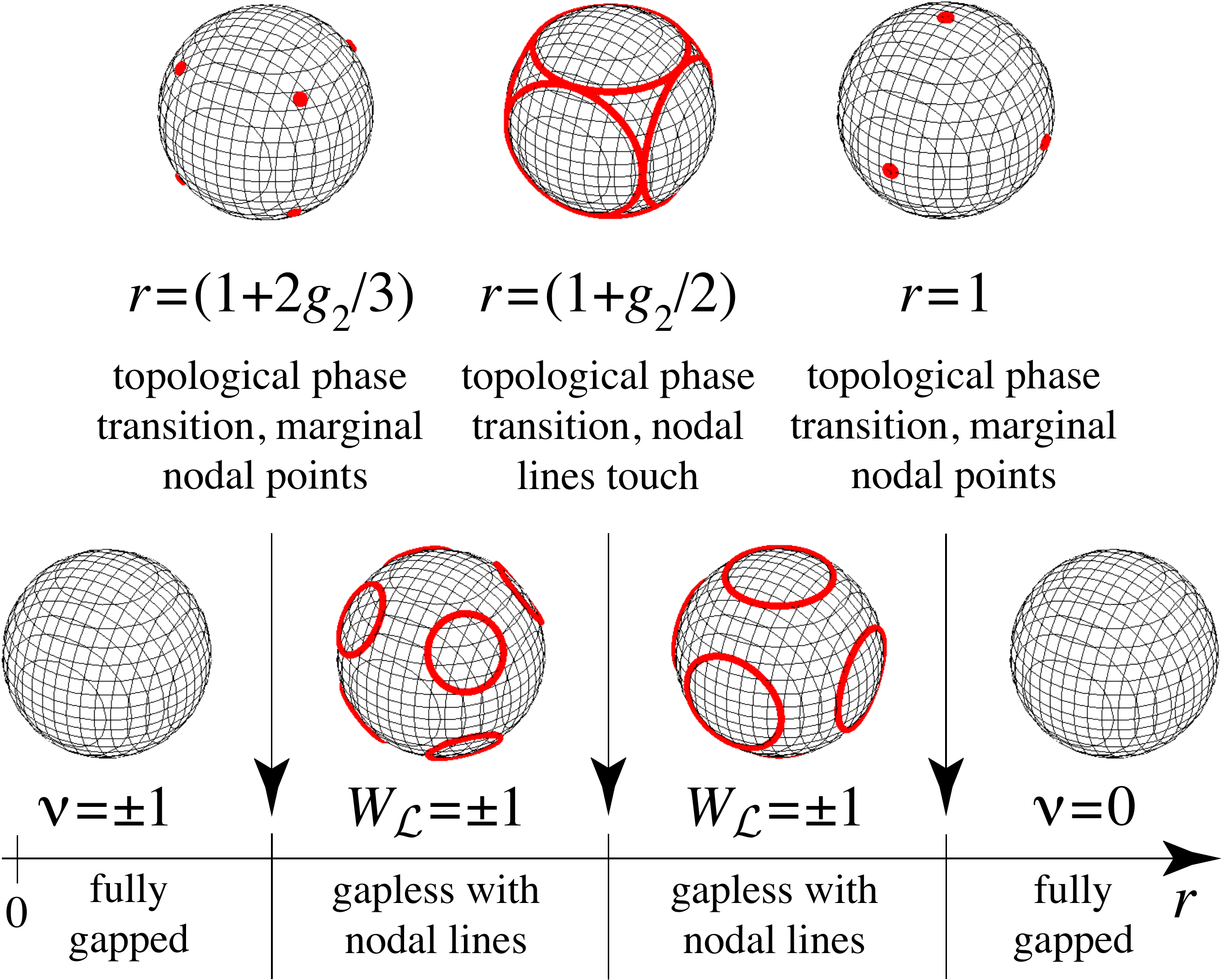}
\caption{\label{Opd}
(Color online)  
Schematic phase diagram for an NCS with cubic point group $O$ as a
function of the ratio $r=\Delta_s / \Delta_t$ of the singlet and triplet gaps. 
Here, the second-order SOC $g_2$ [see Eq.~\eqref{SOC_Ogroup}]
is assumed to be negative.}
\end{figure}

\begin{figure*}[thp]
 \includegraphics[clip,width=2.0\columnwidth]{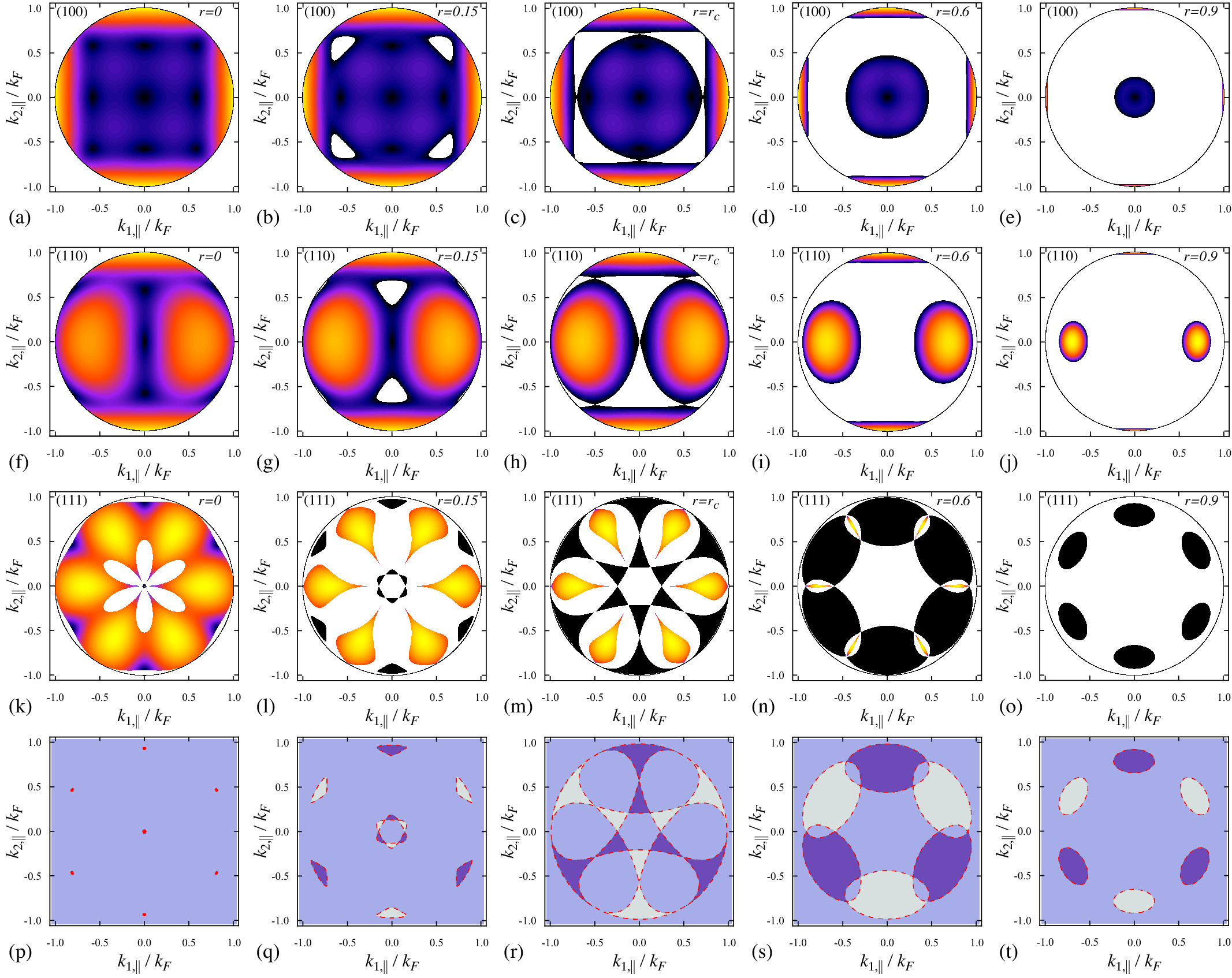}
  \caption{\label{fig3}
  (Color online) 
(a)--(e) Surface bound-state spectra for the (100) face of a NCS with point
  group $O$, $g_2=-1.5$, and $\lambda=0$, as a function of surface momentum
  ${\bf 
  k}_{\parallel}$ for (a) $r = 0$, (b) $r =
  0.15$, (c) $r = r_c = 0.25$, (d) $r = 0.6$, and (e) $r = 0.9$.
  The color scale is the same as in Figs.~\ref{fig1}(a)--(c).
(f)--(j) Same as panels (a)--(e) but for the (110) face.
(k)--(o) Same as panels (a)--(e) but for the (111) face.
(p)--(t) Winding number
  $W_{\textrm{(111)}}$ for the (111) face corresponding to the same parameters
  as in panels (k)--(o). Dark blue (gray) indicates $W_{\textrm{(111)}} = + 1$
  ($-1$), while light blue is $W_{\textrm{(111)}} = 0$.
  The red dashed lines represent the projections of nodal lines.}  
\end{figure*}

\subsection{Surface states for the cubic point group $O$}
\label{sec:bsO}

\begin{figure*}
 \includegraphics[clip,width=2.0\columnwidth]{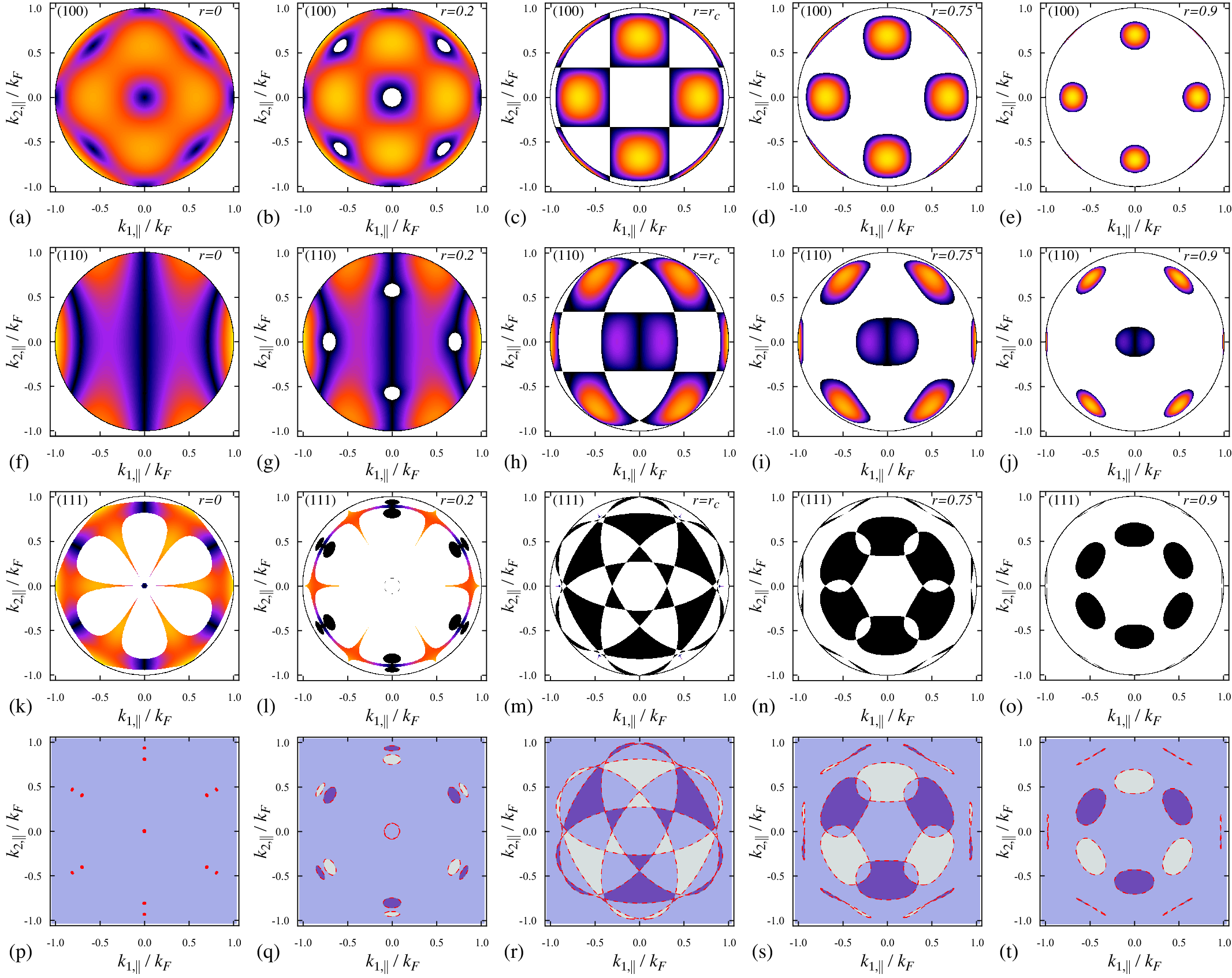}
  \caption{\label{fig4}
  (Color online) 
(a)--(e) Surface bound-state spectra for the (100) face of a NCS with point
  group $T_d$ and $\lambda=0$ as a function of the surface momentum ${\bf k}_{\parallel}$
  for (a) $r = 0$, (b) $r = 0.2$, (c) $r = r_c = 4 \sqrt{2} / 9$, (d) $r =
  0.75$, and (e) $r = 0.9$. The color scale is the same as in
  Figs.~\ref{fig1}(a)--(c).
(f)--(j) Same as panels (a)--(e) but for the (110) face.
(k)--(o) Same as panels (a)--(e) but for the (111) face.
(p)--(t) Winding number $W_{\textrm{(111)}}$ for the (111) face
  corresponding to the same parameters as in panels (k)--(o). Dark blue (gray)
  indicates $W_{\textrm{(111)}} = + 1$ ($-1$), while light blue is
  $W_{\textrm{(111)}} = 0$. The red dashed lines represent the projections
  of nodal lines.}  
\end{figure*}

In the cubic point group $O$ there are pronounced changes in the nodal
structure of the negative-helicity gap as the singlet-to-triplet ratio
$r=\Delta_s / \Delta_t$ is
varied, which are reflected in the surface bound-state spectrum.
To simplify the discussion, we assume a spherical Fermi surface, negligible spin-orbit 
splitting,  and finite $g_2<0$.\cite{g2} 
A schematic topological phase diagram of this NCS is presented in~\fig{Opd}.
For $r>1$ the system is fully
gapped and topologically trivial. Reducing $r$, we find that at $r=1$ point
nodes appear in the negative-helicity gap at 
${\bf k}=k_F(1,0,0)$ and equivalent points.
This is a Lifshitz transition\cite{Lifshitz} at which the topology of the
Bogoliubov-de Gennes
quasiparticle spectrum changes, but the symmetry of the ground state remains 
unaltered.~\cite{volovikTPT,volovikBook1992} 
Upon further lowering $r$,  these point nodes develop into nodal rings
with non-trivial topological charge. A critical value is reached
at $r=r_c=1+g_2/2$, where the nodal rings touch each other and 
reconnect in a different manner, i.e., there is a change in the topology of the
nodal structure itself, with the rings now centered about
${\bf k} = (k_F / \sqrt{3}) ( 1,1,1)$ and equivalent points. Finally,  at $r =
1 + 2 g_2 / 3$ there is another   
Lifshitz phase transition to a fully gapped phase with topologically
non-trivial order characterized by the three-dimensional winding number $\nu =
\pm 1$.\cite{Schnyder2011} The nodal structure at this transition point
  is marginal in the sense that it is topologically trivial and not protected
  against decay either into the fully gapped topologically non-trivial state
  or into the topologically stable gapless phase, i.e. point nodes do not
  posses any topological protection in a three-dimensional NCS. Experimental
  signatures of the topological phase transition will be discussed
  in~\Sec{SevIV}. 
In the following discussion of the bound states we take
$g_2=-1.5$, which implies line nodes for 
$0< r <1$, point nodes for $r=0$, $1$, and a fully gapped state for
$r>1$. Surface bound states only occur for $r < 1$.

We first consider the bound states at the (100) face
[\figs{fig3}(a)--(e)], which were partially 
examined in~\Ref{Vorontsov2008b} for $r>r_c$. Starting at $r=0$ we find bound
states for all $|{\bf k}_\parallel|<k_F$ except at the projected point
nodes. Of particular note are the dispersing zero energy states at
$(k_{1,\parallel},k_{2,\parallel}) = (0,0)$, $(k_F/\sqrt{3},0)$ and
equivalent points, where the condition ${\bf l}_{\bf k} = -{\bf
  l}_{\widetilde{\bf k}}$ is satisfied. From~\Sec{secIII}, however, we
  deduce that only the zero-energy state at the zone center is a
  topologically protected Majorana fermion. The
  other four zero-energy states, in contrast, only form for
  $g_{2}<-1$ and hence are 
  not topologically stable. 
  The apparent zeros close to $(k_F/\sqrt{3},k_F/\sqrt{3})$ and symmetry-related
  points in (a) are the projections of the point nodes.
Upon increasing $r$, the nodal rings on the negative-helicity Fermi
surface lead to regions within the projected Fermi 
surface where bound states do not occur, i.e., the white space
in~\fig{fig3}(b).  
Note that we have $W_{(100)}=0$ everywhere since the integration
path $\mathcal{L}$ passes through none or two nodal rings, the contributions of
which cancel.
The reconnection of the nodal rings at  
$r=r_c$ splits the bound states into disconnected regions at the zone center 
and near the edges of the projected Fermi surface. Note that the Majorana
fermion mode at the zone center survives up to $r=1$. 
As shown in~\figs{fig3}(f)--(j), the situation for the (110) face is
essentially similar, although here the Majorana fermion state
at the zone center only survives up to the topological Lifshitz transition at
$r=r_c$.

\begin{figure*}[t]
  \includegraphics[clip,width=0.667\columnwidth]{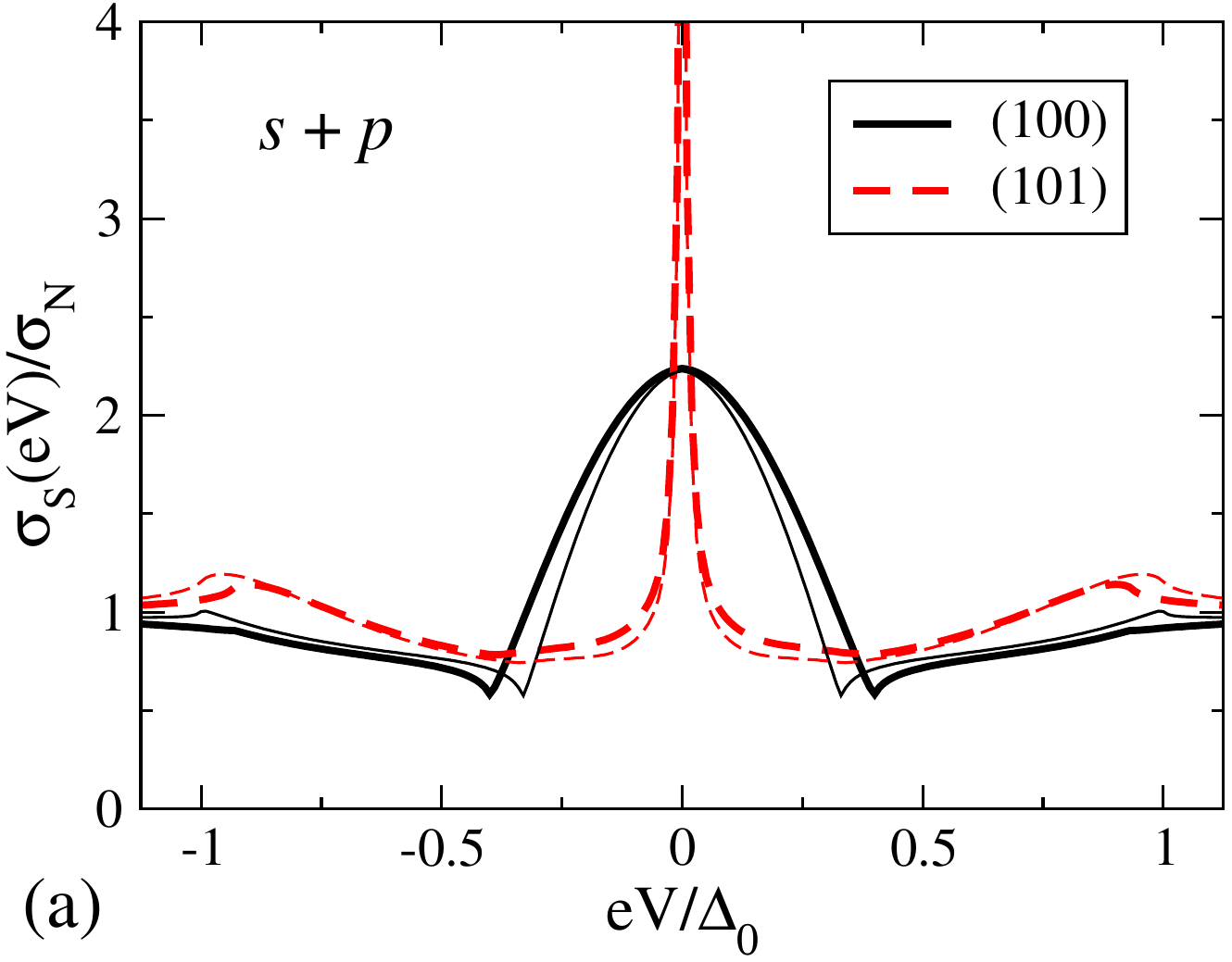}%
  \hspace{0.08cm}\includegraphics[clip,width=0.667\columnwidth]{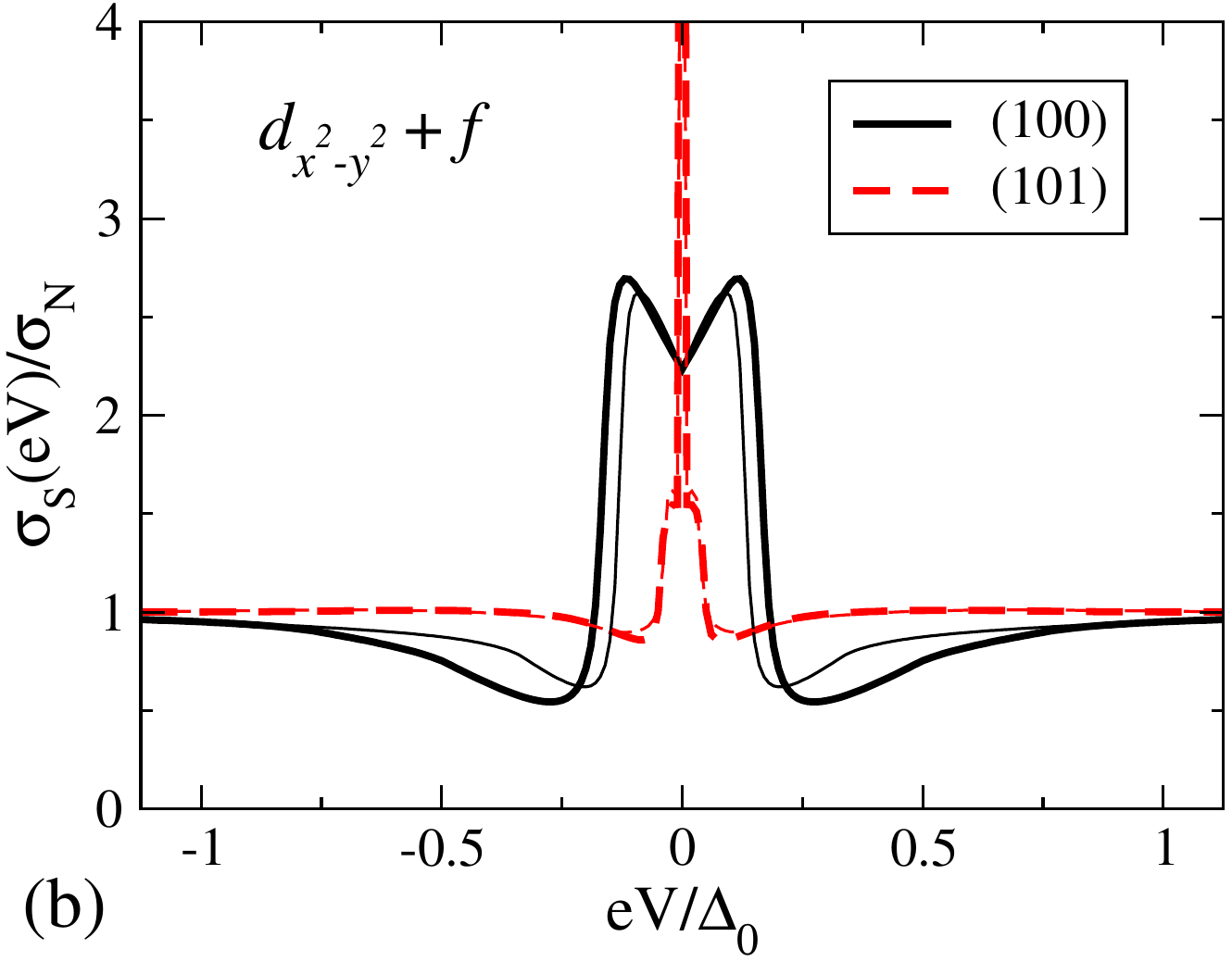}%
  \hspace{0.08cm}\includegraphics[clip,width=0.667\columnwidth]{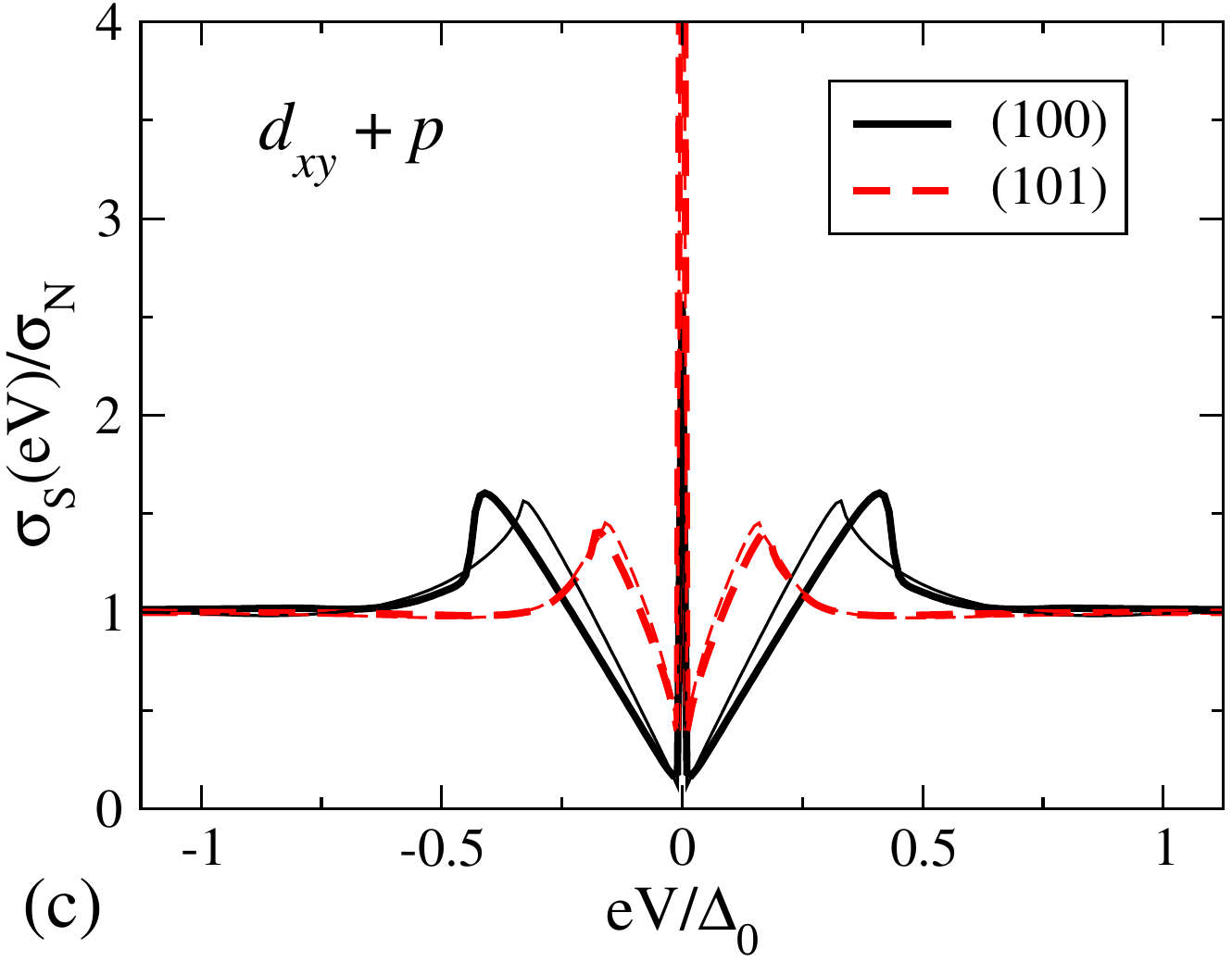}
  \caption{\label{cond}
  (Color online) Tunneling conductance spectra for the  (100) and (101) 
    interfaces of a $C_{4v}$ NCS with $r=0.5$, $Z=3$, and (a) $(s+p)$-wave gap symmetry [$f({\bf
        k})=1$], (b) $(d_{x^2-y^2}+f)$-wave gap symmetry [$f({\bf
        k})=(k_x^2-k_y^2)/k_F^2$], and (c) $(d_{xy}+p)$-wave gap symmetry [$f({\bf
        k})=2k_xk_y/k_F^2$]. In all panels the thick lines show the results
    for finite spin-orbit splitting $\lambda{k_F}/2\mu = 0.1$, while the thin
    lines show the  results for degenerate helical Fermi surfaces.}  
\end{figure*}

The bound states at the (111) face are qualitatively different due to the
presence of topologically protected zero-energy bands. We show the
bound state spectra and the associated maps of the winding number
$W_{(111)}$ 
in~\figs{fig3}(k)--(o) and~\figs{fig3}(p)--(t), respectively. Unlike for the other
two surfaces, we find that at $r=0$ the energy of the dispersing states reaches
the gap so that
we have regions within the projected Fermi surface where bound states do not
occur. For finite but small $r$,
the dispersing bound states break up into six lobes, with zero-energy
flat surface bands within the projected nodal
lines close to the center and edges. Like for the
$C_{4v}$ point group, the zero-energy states are associated with a
winding number $W_{(111)}=\pm1$.
Increasing $r$ we see that 
the zero-energy surface states grow in extent while 
the dispersing states shrink, until at $r=r_c$ the nodes in the gap touch and
the zero-energy states surround the dispersing states. The dispersing
  states vanish at $r$ closer to $1$ when the projected nodes no longer
  overlap, and we hence find only zero-energy states in~\fig{fig3}(o).

\subsection{Surface states for the tetrahedral point group $T_d$}

The tetrahedral point group $T_d$ also shows topological phase
  transitions in the nodal structure upon varying $r$,
which we again discuss for a spherical Fermi surface
and vanishing  spin-orbit splitting. At $r>1$ the system is fully gapped and
topologically trivial. Upon reducing the singlet-to-triplet ratio to $r=1$,
a Lifshitz transition occurs with the appearance of twelve point nodes in the
negative-helicity gap at ${\bf
  k}= (k_F / \sqrt{2} ) (1,1,0)$ and equivalent points; further reducing $r$
we find topologically charged nodal rings centered about these points. These
rings grow as $r$ is lowered, eventually touching at $r=r_c=4\sqrt{2}/9$. For
$r<r_c$ the nodal rings reconnect in a different way so that they are now centered
about the fourteen zeros of ${\bf l}_{\bf k}$ at ${\bf k}=k_F (1,0,0)$, 
$(k_F / \sqrt{3} )(1, 1, 1)$ and equivalent points. In the limit of a
purely triplet gap we find point nodes at these locations.

The bound states at the (100) surface are shown
in~\figs{fig4}(a)--(e). Unlike for
the cubic point group $O$, there are no zero-energy dispersing states for
this surface; the apparent zero-energy states at $r=0$ in~\fig{fig4}(a) are in
fact nodes. For $r>0$ these points become lines, clearly visible as the
boundaries of the white 
space. As $r$ is increased past $r_c$ the dispersing bound states separate
into disconnected regions of the BZ, which vanish at $r=1$.
We now turn to the bound states for the (110) surface, which are presented
in~\figs{fig4}(f)--(j). At $r<r_c$ we find lines of
dispersing zero-energy states which connect the projected point nodes (for $r=0$)
and nodal rings (for $r>0$) of the negative-helicity gap.  These arc states
are not topologically protected by the mechanism discussed in
Sec.~\ref{sec:topCriteria}, however, since for this point-group symmetry 
the condition~\eq{TRSgSec} only 
holds for planes which intersect line nodes of the gap. We nevertheless note
that the zero-energy state at the surface BZ center is a
topologically protected Majorana fermion which is present for all
$r<1$.

Like for the cubic  point group $O$, topologically protected zero-energy bound
states are found at the (111) surface for $0<r<1$. The bound state spectra
are plotted in~\figs{fig4}(k)--(o), with the variation of the
winding number 
$W_{(111)}$ shown in~\figs{fig4}(p)--(t). The main
difference to 
the case of the point group $O$  is the much more limited extent of the
dispersing bound states, which are almost entirely absent for $r>r_c$. 
At small $r$ we also observe arc surface states connecting the
topologically charged nodal rings near the edge of the projected Fermi
surface, but again these states are not topologically protected by
the mechanism discussed in Sec.~\ref{sec:topCriteria}.

\section{Tunneling conductance and topological phase transitions}

In this section we discuss the calculation of the tunneling-conductance
spectra for a normal-metal--NCS junction. Such tunneling experiments are an
important test for the pairing symmetry of unconventional
  superconductors.~\cite{kashiwaya_tanaka00} In particular, they can confirm
  the existence of the zero-energy surface flat bands, which
are evidenced by a sharp peak in the low-temperature conductance at zero
bias.\cite{Brydon2011,Tanaka2010} We also discuss possible signatures of the
topological phase transitions.

\subsection{Tunneling conductance}
\label{sec:tunnelingCond}

At zero temperature the tunneling conductance $\sigma_{S}(eV)$ for tunneling 
from a normal metal without SOC into an NCS under bias
voltage $V$ is given by the generalized Blonder-Tinkham-Klapwijk
formula~\cite{Yokoyama2005,Iniotakis2007,eschrigIniotakisReview,Blonder1982} 
\beq
\sigma_{S}(eV) = \sum_{{\bf k}_\parallel}\sum_{\sigma,\sigma'}\left\{1 +
|a_{\sigma,\sigma'}({{\bf k}_\parallel})|^2 - |b_{\sigma,\sigma'}({{\bf k}_\parallel})|^2\right\} ,
\eeq
where $a_{\sigma,\sigma'}({{\bf k}_\parallel})$ and
$b_{\sigma,\sigma'}({{\bf k}_\parallel})$ are the spin-resolved Andreev
and normal reflection coefficients for electron injection into the NCS,
respectively. These coefficients appear in the wavefunction ansatz describing
the electron-injection process,
\beq
\Psi_{\sigma}({\bf k}_\parallel,{\bf r}) = \psi^{<}_{\sigma}({\bf
  k}_\parallel,{\bf r})\Theta(-r_\perp) + \psi^{>}_{\sigma}({\bf
  k}_\parallel,{\bf r})\Theta(r_\perp)\,.
\eeq
In the normal metal the wavefunction is written as
\beqarray
\psi^{<}_{\sigma}({\bf
  k}_\parallel,{\bf r}) & = & \psi_{e,\sigma}e^{i{\bf k}\cdot{\bf r}} +
\sum_{\sigma'}\left[a_{\sigma,\sigma'}({\bf k}_\parallel)\psi_{h,\sigma'}e^{i{\bf
      k}\cdot{\bf r}} \right. \notag \\
&& \left. + b_{\sigma,\sigma'}({\bf k}_\parallel)\psi_{e,\sigma'}e^{i\widetilde{\bf
      k}\cdot{\bf r}} \right]\, \label{eq:psinormal}
\eeqarray
with the electron and hole spinors $\psi_{e,\sigma} = \frac{1}{2}(1 + \sigma,
1-\sigma, 0,  0 )^{T}$ and $\psi_{h,\sigma}= \frac{1}{2}(0, 0, 1 + \sigma,
1-\sigma)^{T}$, 
respectively.
In the case that there are propagating solutions in both helicity sectors, the
wavefunction in the NCS is written as
\beq
\psi^{>}_{\sigma}({\bf
  k}_\parallel,{\bf r}) = \sum_{n=\pm}\left[c_{\sigma,n}\psi_{n}({\bf
    k}_n)e^{i{\bf k}_n\cdot{\bf r}} + d_{\sigma,n}\psi_{n}(\widetilde{\bf
    k}_n)e^{i\widetilde{\bf k}_n\cdot{\bf r}}\right] , \label{eq:psiNCS}
\eeq
where the spinors are defined as in~\Sec{sec:2prop}; if there
are propagating solutions only in the negative-helicity sector, the
positive-helicity components of~\eq{eq:psiNCS} are replaced by~\eq{eq:Psicase2}. 
To simplify the discussion we assume that the normal metal has a spherical
Fermi surface with the same
chemical potential and effective mass as the NCS. We restrict
ourselves to the limit where the bias energy is negligible compared to the
Fermi energy so that the energy dependence of the wavevectors can be ignored.

The reflection coefficients are determined by application
of the interface boundary conditions. The first condition requires continuity
of the wavefunction across the interface, i.e.,
\beq
\psi^{<}_{\sigma}({\bf k}_\parallel,{\bf r})|_{r_\perp=0^-}  =
\psi^{>}_{\sigma}({\bf k}_\parallel,{\bf r})|_{r_\perp=0^+} \, .
\eeq
The second condition enforces conservation of probability across the
barrier,~\cite{Molenkamp2001,Zuelicke2001,Matsuyama2002} which is modeled as
a $\delta$-function of height $U$. We have
\beq
\check{v}_{S}\psi^{>}_\sigma({\bf
  k}_\parallel,{\bf r})|_{r_\perp=0^{+}} =
\left(\partial_{r_\perp} + 2Zk_F\right)\psi^{<}_\sigma({\bf 
  k}_\parallel,{\bf r})|_{r_\perp=0^{-}}\, ,
\eeq
where $Z=mU/\hbar^2k_F$ is a dimensionless parameter characterizing the
transparency of 
the interface, and $\check{v}_S$ is proportional to the
$r_\perp$ component of the velocity operator in the normal state of the NCS.  
For a $C_{4v}$ point-group NCS orientated such that the $b$ axis is
parallel to the 
interface and the $c$ axis makes an angle $\alpha$ to the interface, we write 
\beq
\check{v}_S =  \hat{\sigma}_{0}\otimes\hat{\sigma}_0\,\partial_{r_\perp} + 
i\,\hat{\sigma}_0\otimes\hat{\sigma}_2\,\frac{m\lambda}{\hbar^2}\,\cos\alpha \, .
\eeq

In~\fig{cond} we present conductance spectra for  tunneling
  into a $C_{4v}$ point-group NCS through
(100) and (101) interfaces for both vanishing
and finite SOC 
$\lambda$. We find the tunneling-conductance results to be quite 
robust against the finite spin-orbit splitting of the Fermi surfaces. 
Several features of these spectra are noteworthy. For the
(100) surface we observe a broad hump-like feature in the tunneling
conductance for the ($s+p$)-wave and ($d_{x^2-y^2}+f$)-wave pairing states,
which is a signature of the arc surface states. In the ($d_{xy}+p$)-wave case,
in contrast, we find a zero-bias conductance peak well-separated from the bulk
density of states. In the limit of degenerate
helical Fermi surfaces this is due to the
doubly degenerate zero-energy states for ${\bf k}_{\parallel}$ close to $\pm
k_{F}{\bf e}_{z}$ [see~\fig{fig1}(d)], while for non-zero SOC
the singly degenerate zero-energy states near ${\bf k}_\parallel=\pm k_F{\bf
  e}_y$ also contribute. 
For the (101) surface all pairing symmetries show a zero-bias conductance
peak, which is a key experimental signature of the topologically
protected zero-energy states.~\cite{Brydon2011,Tanaka2010} For the cases of
($s+p$)-wave and ($d_{x^2-y^2}+f$)-wave pairing, however, we note that this is
superimposed on a
much diminished hump-like feature. This signals the continued existence of arc
surface states in these systems, in agreement with~\fig{fig2}.

\subsection{Evidence for topological phase transitions}
\label{SevIV}

\begin{figure}[t!]
\includegraphics[clip,width=0.9\columnwidth]{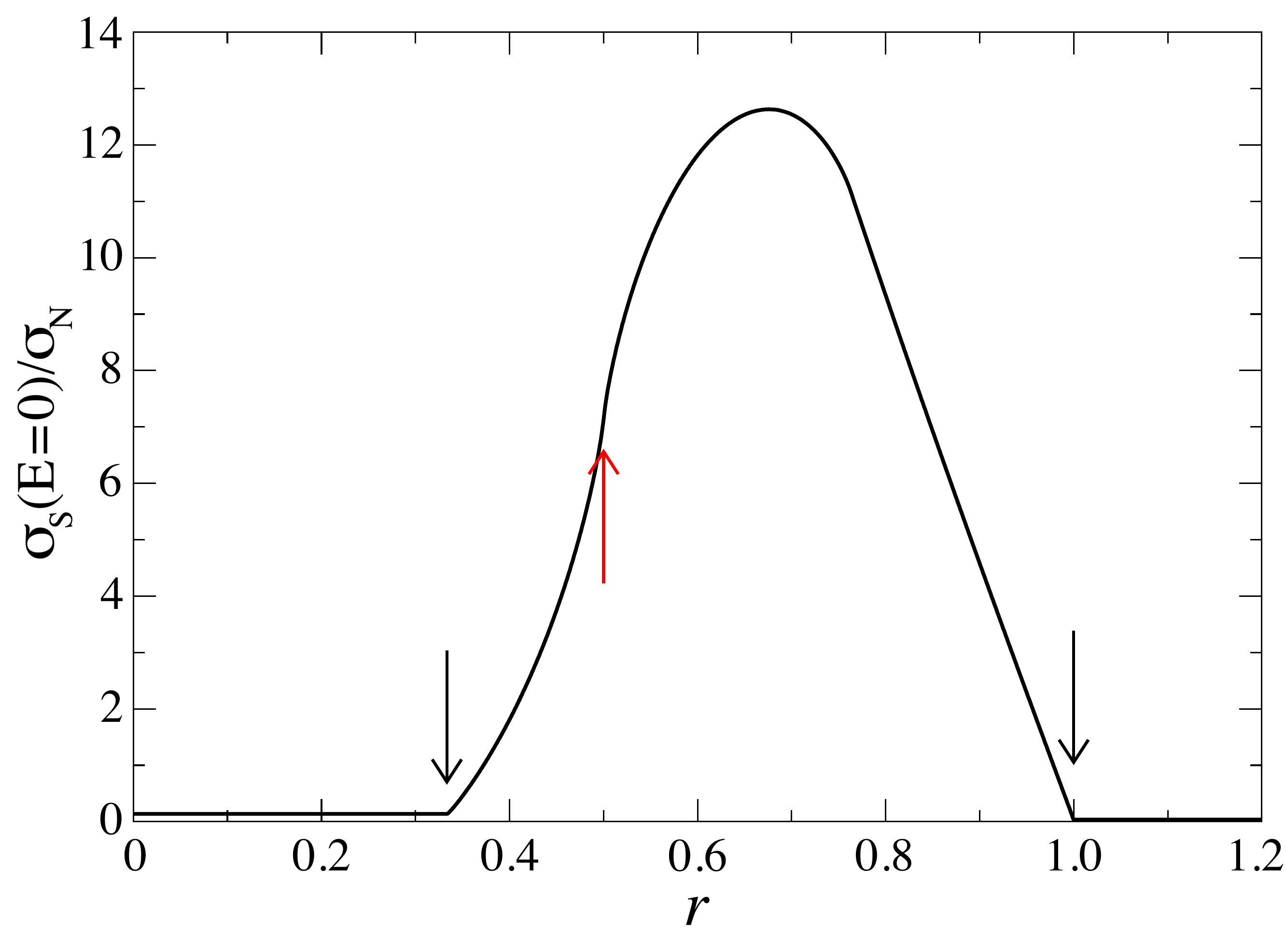}
\caption{\label{zbcpvsq}
(Color online)  
Variation of the zero-bias conductance-peak
height at the (111) face as a function of singlet-to-triplet ratio $r$ 
for the point group $O$. The black, downward-pointing arrows indicate topological phase
transitions between fully gapped and nodal phases.
The red, upward-pointing arrow marks an inflection point  in $\sigma_{S}(eV)$, where
the character of the nodes on $\Delta^{-}_{\bf k}$ changes qualitatively.
In this figure we set $g_2=-1$, $Z=3$, and $T=0\,\mathrm{K}$.}
\end{figure}

In~\Sec{sec:bsO} we have demonstrated the existence of topological phase
transitions as a function of the singlet-triplet ratio $r$ for an NCS with
point group $O$ and SOC vector ${\bf l}_{\bf k}$ given by
Eq.~\eqref{SOC_Ogroup} with 
$g_{2}\neq0$. This discussion is physically relevant for
Li$_2$Pd$_x$Pt$_{3-x}$B, for which the SOC strength can be tuned by
substituting Pd for Pt.\cite{leePickett2005} 
The magnitude of the SOC interaction in these compounds seems to be
directly related to $r$,\cite{yuan2006}  which
suggests that it might be possible to observe  
topological phase transitions between a fully gapped and a gapless phase, or
between two gapless phases, in Li$_2$Pd$_x$Pt$_{3-x}$B
as a function of Pt concentration.

The most direct way to detect the topological phase transitions in this system
requires measurements sensitive to the low-energy bulk density of states
$\rho(\omega)$. At the $r=1$ and $r = ( 1 + 2 g_2 / 3 )$ Lifshitz transitions,
this quantity changes from 
$\rho(\omega ) = 0$, characteristic of a full gap, to the linear dependence
$\rho (\omega) \propto \omega$ associated with line nodes. Signatures of the
topological phase transition can also be seen in the conductance spectra, in
particular the zero-bias conductance peak at the (111)
surface. Indeed, as shown in~\fig{zbcpvsq}, the zero-bias conductance
shows abrupt changes at the boundaries of the nodal
region. Furthermore, at the critical $r_c$ where the topological structure of
the nodes changes, there is a kink anomaly which is marked by the red,
upward-pointing arrow
in~\fig{zbcpvsq}. Tunneling experiments could therefore in principle be used
to evidence a topological phase transition in Li$_2$Pd$_x$Pt$_{3-x}$B.

\section{Conclusions and Outlook}
\label{sec:conclusions}

We have performed a detailed analysis of the topological
properties of nodal NCSs in three dimensions. Using topological arguments, we
have derived general criteria 
for the existence of Andreev bound states at the surface of nodal NCSs. Three
different types of topologically protected surface states have been
identified, namely Kramers-degenerate Majorana modes, arc surface states, and
surface flat bands, whose stability is guaranteed by the conservation of the
one- and two-dimensional $\mathbbm{Z}_2$ invariants  [Eq.~\eqref{1dZ2first}
  and \eqref{Z2noLattice}], and the winding number~\eqref{eq: 1D winding no},
respectively.

We have independently derived the general criteria for zero-energy surface
states using the quasiclassical scattering theory. Furthermore, we have
applied this technique to  
study a number of physically relevant manifestations of NCSs. For a $C_{4v}$
point-group symmetry, we have calculated the surface bound-state spectra and the
tunneling conductance for ($s+p$)-wave, ($d_{x^2-y^2}+f$)-wave, and
$(d_{xy}+p$)-wave pairing and finite spin-orbit splitting. We have shown that
the surface bound states are in perfect agreement with the variation of the
topological winding number across the surface BZ. We have  
shown how the surface flat bands manifest themselves as a zero-bias
conductance peak, while the arc surface states lead to a broad, hump-like
feature centered around zero bias in the conductance spectra. Both
features exhibit a pronounced
dependence on surface orientation, which provides
characteristic fingerprints of the topological properties of the system.
We have also examined the bound-state spectra in NCSs with $O$ and $T_d$
point-group symmetry in the limit of weak spin-orbit splitting and discussed the
occurrence of topological Lifshitz transitions in these spectra.

On symmetry grounds the
zero-energy surface flat bands are expected to appear in any unconventional
nodal superconductor preserving TRS. In particular,
this should be the case for the class-CI topological
superconductors.\cite{schnyderPRL09,schnyderPRB10,hosur2010} Furthermore, flat
bands are also expected to occur at the surface of nodal topological
superconductors in symmetry class AIII\cite{Schnyder2011,hosur2010} 
(i.e., topological superconductors with TRS that are
invariant under rotations about one fixed axis in spin space). 
Besides the tunneling conductance, the surface flat bands and arc surface
states also profoundly affect other surface and interface properties of
NCSs, such as Josephson tunneling,\cite{Asano2011} the non-linear Meissner effect,
and surface
thermal transport. The investigation of these interesting boundary properties
are left for future work.

\acknowledgments 

The authors thank L.\ Klam, S.\ Ryu,  M.\ Sigrist, and G.\ Khaliullin for
useful discussions.

\end{document}